\documentclass[twocolumn]{jpsj3}

\def\runauthor{M. Koga, M. Matsumoto, and H. Kusunose}
\def\runauthor{}

\title{
Effect of Impurities with Internal Structure on Multiband Superconductors \\
-- Possible Enhancement of Transition Temperature --
}

\author{Mikito Koga, Masashige Matsumoto$^1$, and Hiroaki Kusunose$^2$}

\inst{
Department of Physics, Faculty of Education, Shizuoka University, 836 Oya, Suruga-ku, Shizuoka
422--8529 \\
$^1$Department of Physics, Faculty of Science, Shizuoka University, 836 Oya, Suruga-ku, Shizuoka
422--8529 \\
$^2$Department of Physics, Ehime University, Matsuyama 790-8577
}

\recdate{\today}

\abst{
We study inelastic (dynamical) impurity scattering effects in two-band superconductors with the same
($s_{++}$ wave) or different ($s_\pm$ wave) sign order parameters.
We focus on the enhancement of the superconducting transition temperature $T_{\rm c}$ by magnetic
interband scattering with the interchange of crystal-field singlet ground and multiplet excited states.
Either the $s_{++}$-wave or $s_\pm$-wave state is favored by the impurity-mediated pairing,
which depends on the magnetic and nonmagnetic scattering strengths derived from the hybridization
of the impurity states with the conduction bands.
The details are examined for the singlet-triplet configuration that is suggestive of Pr impurities
in the skutterudite superconductor LaOs$_4$Sb$_{12}$.
}

\kword{
multiband superconductivity, $s_{++}$-wave superconductivity, $s_\pm$-wave superconductivity,
heavy fermion, skutterudite, impurity, crystal field
}

\begin{document}

\maketitle

\newcommand{\ds}{\displaystyle}

\renewcommand{\H}{\mathcal{H}}
\newcommand{\br}{{\mbox{\boldmath$r$}}}
\newcommand{\bR}{{\mbox{\boldmath$R$}}}
\newcommand{\bM}{{\mbox{\boldmath$M$}}}
\newcommand{\bS}{{\mbox{\boldmath$S$}}}
\newcommand{\bk}{{\mbox{\boldmath$k$}}}
\newcommand{\bPsi}{{\mbox{\boldmath$\Psi$}}}
\newcommand{\bpsi}{{\mbox{\boldmath$\psi$}}}
\newcommand{\bPhi}{{\mbox{\boldmath$\Phi$}}}
\newcommand{\bG}{{\hat{G}}}
\newcommand{\om}{{\omega_l}}
\newcommand{\omd}{{\omega^2_l}}
\newcommand{\btau}{{\hat{\tau}}}
\newcommand{\brho}{{\hat{\rho}}}
\newcommand{\bsigma}{{\hat{\sigma}}}
\newcommand{\bSigma}{{\hat{\Sigma}}}
\newcommand{\bU}{{\hat{U}}}
\newcommand{\bT}{{\hat{T}}}
\newcommand{\bA}{{\hat{A}}}
\newcommand{\bB}{{\hat{B}}}
\newcommand{\bP}{{\hat{P}}}
\newcommand{\blambda}{{\hat{\lambda}}}
\newcommand{\bDelta}{{\hat{\Delta}}}
\newcommand{\bLambda}{{\hat{\Lambda}}}
\newcommand{\bV}{{\hat{V}}}
\newcommand{\bskp}{{\mbox{\scriptsize\boldmath $k$}}}
\newcommand{\skp}{{\mbox{\scriptsize $k$}}}
\newcommand{\bsQp}{{\mbox{\scriptsize\boldmath $Q$}}}
\newcommand{\bsqp}{{\mbox{\scriptsize\boldmath $q$}}}
\newcommand{\bsrp}{{\mbox{\scriptsize\boldmath $r$}}}
\newcommand{\bsRp}{{\mbox{\scriptsize\boldmath $R$}}}
\newcommand{\bsk}{\bskp}
\newcommand{\sk}{\skp}
\newcommand{\bsQ}{\bsQp}
\newcommand{\bsq}{\bsqp}
\newcommand{\bsr}{\bsrp}
\newcommand{\bsR}{\bsRp}
\newcommand{\bbsigma}{{\mbox{\boldmath$\hat{\sigma}$}}}
\newcommand{\ri}{{\rm i}}
\newcommand{\re}{{\rm e}}
\newcommand{\rd}{{\rm d}}
\newcommand{\Tc}{{$T_{\rm c}$}}
\renewcommand{\Pr}{{PrOs$_4$Sb$_{12}$}}
\newcommand{\La}{{LaOs$_4$Sb$_{12}$}}
\newcommand{\LaPr}{{La$_{1-x}$Pr${_x}$Os$_4$Sb$_{12}$}}

\section{Introduction}

The study of impurity effects on multiband superconductivity is stimulated by the discovery
of high-temperature (high \Tc) superconductors with FeAs layers,
\cite{Kamihara08,Takahashi08,Kawabata08,Sefat08,Leithe08,Lee09,Sato09,Ishida09}
while it has been performed for another high-\Tc~superconductor, MgB$_2$.
\cite{Nagamatsu01,Kortus05,Putti07}
One of the interesting points for the impurity problem is their anisotropic band property.
In particular, much attention is paid to the $s_\pm$-wave pairing state characterized by sign
reversal of order parameters with full gaps.
\cite{Mazin08,Kuroki08,Chubukov08,Parker08,Bang09}
For a depairing effect caused by impurity scattering, the $s_\pm$-wave state behaves like $d$-wave
pairing rather than the conventional $s$-wave state.
\cite{Golubov97}
Multiband properties are also reported in the heavy-fermion superconductor \Pr~whose higher
\Tc~than \La~implies a crucial role of the Pr $f$-electron states.
\cite{Seyfarth05,Yogi06}
Interband scattering due to impurities may be important in the multiband, the roles of which
remain to be elucidated.
\par

Recently, Senga and Kontani have studied the effects of intraband and interband nonmagnetic impurity
scatterings in the $s_\pm$-wave state using a simple two-band BCS model.
\cite{Senga08,Senga09}
They found that \Tc~is not markedly suppressed when the intraband and interband scattering
strengths are not equal and are sufficiently strong.
We also applied a similar model in a magnetic impurity case and solved a single-impurity
problem.
\cite{Matsu09}
For the interband scattering in the $s_\pm$-wave state, the roles of magnetic impurities are equivalent
to those of nonmagnetic impurities for the intraband scattering.
\cite{Golubov97}
Very recently, it has been reported that \Tc~is not markedly suppressed so much by the interband
magnetic scattering.
\cite{Li09}
These results imply a possibility of \Tc~enhancement due to magnetic impurities having an internal
structure.
Such enhancement was argued by Fulde {\it et al.} for nonmagnetic impurities in single-band
superconductors.
\cite{Fulde70}
An attractive interaction stems from inelastic nonmagnetic scattering
by impurities, which is analogous to the electron-phonon origin of BCS superconductors. 
\Tc~can be enhanced by doping such impurities with appropriate crystal-field level splitting,
while it is suppressed by magnetic impurities.
\cite{Fulde70}
This idea has recently been applied to a skutterudite superconductor \LaPr~to account for the
\Tc~enhancement by Pr substitution for La in \La .
\cite{Chang07}
Since only a single band is considered there, one always finds that magnetic impurities cause
\Tc~suppression.
However, this conventional understanding has to be reexamined for multiband superconductors.
In the case of $s_\pm$-wave superconductivity, magnetic impurities for intraband scattering
behave like nonmagnetic impurities for interband scattering, and vice versa.
Accordingly, we can expect \Tc~to be enhanced by the inelastic (dynamical) magnetic scattering.
\par

In this paper, we extend the theory of Fulde {\it et al.}
\cite{Fulde70}
and show a case of magnetic impurities in the $s_\pm$-wave state in \S2.
Considering spin-dependent interband scattering, we examine a self-energy in the Born approximation
to derive a gap equation for two bands.
The increase in \Tc~is associated with the sign change of the superconducting order parameter via the inelastic impurity scattering.
It is demonstrated for the $s_\pm$-wave state whose \Tc~is enhanced by the magnetic interband scattering due to the singlet-singlet configuration.
In \S3, we also show some typical examples of the corresponding interband scattering.
There are various octupolar (combination of spin and orbital) scattering types that increase
\Tc~accompanied by an interchange between two crystal-field singlet states.
The same argument is applied to the singlet-multiplet configuration that can be realized for such $f$-electron impurities
as rare-earth or actinide ions embedded in cubic or uniaxial anisotropic (tetragonal or hexagonal) crystals.
For the singlet-doublet, we discuss the $s_\pm$-wave and $s_{++}$-wave combination via intraband scattering.
The $s_{++}$-wave pairing is characterized by the same sign order parameters of the two bands.
We show a possible crystal-field splitting for \Tc~enhancement, however, \Tc~is suppressed by the
competition between magnetic and nonmagnetic scattering effects.
For the singlet-triplet, we find a case in which the $s_{++}$ wave is favorable for the \Tc~enhancement
by magnetic interband scattering.
Which is chosen, the $s_\pm$ wave or the $s_{++}$ wave, by the impurity-mediated pairing depends
on (1) the scattering type (dipole, quadrupole, octupole, etc.) and (2)the hybridization between the
impurity atomic orbitals and conduction bands.
The details are described for an impurity with the singlet-triplet configuration.
Its application to Pr impurity effects in the \La~superconductor is discussed in \S4.
Conclusions are given in \S5.
\par

\section{Formulation for \Tc~Enhancement}

In this section, we present a formulation for \Tc~enhancement by inelastic impurity scattering in
multiband superconductors.
First, we briefly review a work by Fulde {\it et al}. for single-band superconductors.
\cite{Fulde70}
Then, we extend it to a two-band system as a simple case of multiband and give an example to understand multiband effects on \Tc.
\par

\subsection{Inelastic impurity scattering and \Tc~enhancement in single-band superconductors}

Let us begin with the model Hamiltonian $\H = \H_{\rm C} + \H_{\rm I} + \H'$ that consists of conduction electron $\H_{\rm C}$,
impurity $\H_{\rm I}$, and impurity scattering $\H'$ terms.
The first term is written as
\begin{align}
&\H_{\rm C} = \sum_\sigma \int\rd\br \psi_{\sigma}^\dagger(\br) \epsilon (-\ri\nabla) \psi_{\sigma}(\br) \cr
&~~~~~~
            - \Delta \int\rd\br \left[ \psi_{\uparrow}^\dagger(\br)\psi_{\downarrow}^\dagger(\br)
                                     + \psi_{\downarrow}(\br)\psi_{\uparrow}(\br) \right].
\end{align}
Here, $\psi_{\sigma}(\br)$ is a field operator of the conduction election of the
$\sigma(=\uparrow,\downarrow)$ spin
whose kinetic energy is expressed by the operator $\epsilon(\ri\nabla)=-\nabla^2/2m_{\rm e} -E_{\rm F}$ measured from the Fermi energy $E_{\rm F}$,
where $m_{\rm e}$ represents the electron mass and the Planck constant $\hbar$ is taken as unity.
$\Delta$ is the $s$-wave superconducting order parameter that we assume to be a real value.
The Hamiltonian $\H_{\rm I}$ for the impurity states is given by
\begin{align}
\H_{\rm I} = \sum_{\bsR_\gamma} \sum_{m} \delta_m a_{\gamma m}^\dagger a_{\gamma m}.
\end{align}
Here, $\bR_\gamma$ represents the position of the $\gamma$th impurity.
$a_{\gamma m}^\dagger$ and $a_{\gamma m}$ are the pseudo-fermion creation and annihilation
operators, respectively, for the $m$th impurity energy level $\delta_m$ at the $\gamma$th impurity site.
\cite{Abrikosov65}
The interaction Hamiltonian at the impurities is defined by
\begin{align}
\H' = \sum_{\bsR_\gamma} \sum_{mn} \sum_{\sigma} \int\rd\br
      a_{\gamma m}^\dagger a_{\gamma n} \delta(\br-\bR_\gamma) M_{mn} \psi_{\sigma}^\dagger(\br) \psi_{\sigma}(\br).
\end{align}
We consider here only nonmagnetic impurity scattering since magnetic impurity scattering cannot
enhance \Tc~in a single-band case.
\cite{Fulde70}
$M_{mn}$ is a matrix element that describes the scattering of conduction electrons
accompanied by an interchange among the $m$th and $n$th energy levels.
\par

\begin{figure}[t]
\begin{center}
\includegraphics[width=6cm]{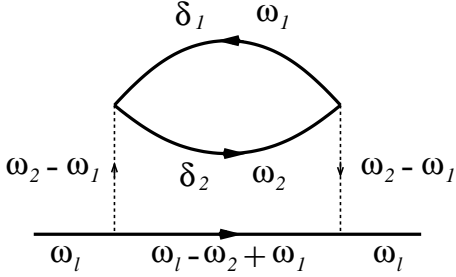}
\end{center}
\caption{
Feynman diagram of self-energy in the second Born approximation.
}
\label{fig:selfenergy}
\end{figure}

We introduce the following $4\times 4$ matrix form of the thermal Green's function:
\begin{align}
\bG(\tau,\br,\br') = - \langle T \bPsi(\br,\tau) \bPsi^\dagger(\br',0) \rangle,
\end{align}
where $\bPsi(\br)$ and $\bPsi^\dagger(\br)$ are four-dimensional vectors defined as
\begin{align}
& \bPsi (\br) =
   \left(
     \begin{array}{c}
       \psi_{\uparrow}(\br) \\
       \psi_{\downarrow}(\br) \\
       \psi_{\uparrow}^\dagger(\br) \\
       \psi_{\downarrow}^\dagger(\br)
     \end{array}
   \right), \cr
& \bPsi^\dagger(\br) =
   \left(
     \begin{array}{cccc}
       \psi_{\uparrow}^\dagger(\br) &  \psi_{\downarrow}^\dagger(\br) & \psi_{\uparrow}(\br) & \psi_{\downarrow}(\br)
     \end{array}
   \right),
\end{align}
with their Heisenberg representations
\begin{align}
\bPsi(\br,\tau) = \re^{\H\tau} \bPsi(\br) \re^{-\H\tau},~~~~~~
\bPsi^\dagger(\br,\tau) = \re^{\H\tau} \bPsi^\dagger(\br) \re^{-\H\tau}.
\label{eqn:Heisenberg}
\end{align}
In the absence of impurity scattering, the unperturbed Green's function is Fourier-transformed to
\begin{align}
\bG_0(\ri\om,\bk)
  = - \frac{ \ri\om + \epsilon_{\bsk}\brho_3 + \Delta\brho_2 \bsigma_2 } { \omd + \epsilon_{\bsk}^2 + \Delta^2 },
\label{eqn:G88-0}
\end{align}
where $\bsigma_\alpha$ and $\brho_\alpha$ ($\alpha$ is denoted by $1$, $2$, and $3$
instead of $x$, $y$, and $z$, respectively, hereafter)
are the Pauli matrices for the spin space and particle-hole space, respectively.
Similarly, the matrix for impurity scattering is given by
\begin{align}
\bU_{mn} = M_{mn} \brho_3.
\label{eqn:U0-matrix}
\end{align}
Following Fulde {\it et al.},
\cite{Fulde70}
we study the \Tc~enhancement on the basis of the second Born approximation.
For the scattering matrix $\bU_{mn}$ in eq.~(\ref{eqn:U0-matrix}),
Fig.~\ref{fig:selfenergy} shows the self-energy given by
\begin{align}
& \bSigma(\ri\om) = - n_{\rm imp}T^2 \sum_{mn}
  \sum_{\omega_1\omega_2} \frac{1}{\ri\omega_1-\delta_m} \frac{1}{\ri\omega_2-\delta_n} \cr
&  ~~~~~~\times
  \frac{1}{\Omega} \sum_\bsk \bU_{mn} \bG_0(\ri\om+\ri\omega_1-\ri\omega_2,\bk) \bU_{nm}.
\end{align}
Here, $n_{\rm imp}$ represents the impurity density.
$1/(\ri\omega_1-\delta_m)$ is an unperturbed Green's function for the $m$th impurity energy level.
$\omega_1$ and $\omega_2$ are Matsubara frequencies for fermions.
$\Omega$ represents the system volume.
The Boltzmann constant $k_{\rm B}$ is taken as unity.
The important point is that $\Delta$ in eq.~(\ref{eqn:G88-0}) changes its sign with the
$\bU_{mn} \bG_0\bU_{nm}$
transformation.
This is a key to the \Tc~enhancement due to the inelastic impurity scattering,
by analogy with the optical phonon case leading to an attractive interaction for pairing.
\cite{Fulde70}
On the other hand, for the elastic impurity scattering, the pairing interaction is cancelled out by
a depairing effect.
\par

In the presence of the impurities, the linearized gap equation is given by
\begin{align}
\Delta \log \frac{ T_{\rm c} }{ T_{{\rm c} 0} } = \pi T_{\rm c} \sum_l \frac{\Delta}{ | \om | }
\left[ \Sigma_\Delta ( \ri \om ) - \ri \frac{ \Sigma_\omega ( \ri \om ) }{ \om } \right],
\label{eqn:single-gap}
\end{align}
where $T_{\rm c}$ ($T_{{\rm c} 0}$) is the transition temperature in the presence (absence) of impurities.
$\Sigma_\Delta$ and $\Sigma_\omega$ are self-energies corresponding to the order parameter and Matsubara frequency components, respectively.
\cite{Fulde70}
In Green's function, the renormalized frequency $\tilde{\om}$ and the order parameter $\tilde{\Delta}_l$
are given by
\begin{align}
\tilde{\om} = \om + \ri \Sigma_\omega ( \ri \om ),~~~~~~
\tilde{\Delta}_l = \Delta [ 1 + \Sigma_\Delta ( \ri \om ) ].
\end{align}
The self-energy in Fig.~\ref{fig:selfenergy} leads to different signs of $\Sigma_\Delta$ in eq.~(\ref{eqn:single-gap})
between time-reversal invariant scattering and non-time-reversal invariant scattering.
\cite{Fulde70}
The former (nonmagnetic scattering in a single band) contributes to \Tc~enhancement.
In a simple case of two energy levels ($\delta_2>\delta_1$), after the lengthy calculation of the
self-energy, we obtain each term on the right-hand side of eq.~(\ref{eqn:single-gap}) as
\begin{align}
& \pi T_{\rm c} \sum_l \frac{1}{| \om |} \Sigma_\Delta (\ri \om)
= - \frac{\pi}{8 T_{\rm c} \tau_{12}} f_\Delta (x), \cr
&~~~~~~ f_\Delta (x) = - \frac{\tanh x}{x} + A(x) - \frac{1}{2} B(x), \\
& \pi T_{\rm c} \sum_l \frac{\ri}{| \om |} \frac{\Sigma_\omega (\ri \om)}{\om}
= - \frac{\pi}{8 T_{\rm c} \tau_{12}} f_\omega (x), \cr
&~~~~~~ f_\omega (x) = - 1 + \tanh^2{x} - \frac{1}{2} B(x), \cr
&~~~~~~ \frac{1}{\tau_{12}} = 2\pi n_{\rm imp}N_0 |M_{12}|^2,
\nonumber
\end{align}
with
\begin{align}
& x = \frac{\delta_2-\delta_1}{2T_{\rm c}} > 0,~~A(x) = S_1(x) \tanh x,~~B(x) = S_2(x) \tanh x,
\label{eqn:x} \\
&~~~~~~ S_1(x) = \frac{4x}{\pi^4} {\rm Re} \sum_{n=0}^\infty
\frac{\ds \psi\left( 1+n-\ri{\frac{x}{\pi}} \right) - \psi\left({\frac{1}{2}}\right)}
{\ds \left(n+\frac{1}{2}\right) \left(n+\frac{1}{2}-\ri\frac{x}{\pi}\right)^2}, \cr
&~~~~~~ S_2(x) = \frac{8}{\pi^3} {\rm Im} \sum_{n=0}^\infty
\frac{\ds \psi\left( 1+n-\ri\frac{x}{\pi} \right) - \psi\left(\frac{1}{2}\right)}
{\ds \left(n+\frac{1}{2}-\ri\frac{x}{\pi}\right)^2}.
\nonumber
\end{align}
Here, $\tau_{12}$ represents the lifetime due to the impurity scattering.
$N_0$ is the density of states at the Fermi energy.
$\psi(x)$ is the digamma function.
As a consequence, eq.~(\ref{eqn:single-gap}) leads to a gap equation for the nonmagnetic scattering
($\tau_{12}$ is denoted by $\tau_{12}^0$),
\begin{align}
\frac{8 T_{\rm c}}{\pi} \tau_{12}^0 \log \frac{T_{\rm c}}{T_{{\rm c}0}}
= - f_\Delta (x) + f_\omega (x) \equiv f_0 (x) > 0,
\label{eqn:f0}
\end{align}
and to that for the magnetic scattering ($\tau_{12}$ is denoted by $\tau_{12}^{\rm s}$),
\begin{align}
\frac{8 T_{\rm c}}{\pi} \tau_{12}^{\rm s} \log \frac{T_{\rm c}}{T_{{\rm c}0}}
= f_\Delta (x) + f_\omega (x) \equiv - f_s (x) < 0.
\label{eqn:fs}
\end{align}
These equations show that for any finite crystal-field splitting, \Tc~is enhanced by nonmagnetic
impurities in single-band $s$-wave superconductors, while it is suppressed by magnetic impurities.
\cite{Fulde70}
For the nonmagnetic scattering in eq.~(\ref{eqn:f0}), we note that \Tc~keeps $T_{{\rm c}0}$ for $x=0$
(elastic scattering) that gives $f_0 (0) = 0$ [$f_\Delta (0) = f_\omega (0) = -1$], while
$|f_\Delta (x)| > |f_\omega (x)|$ [$f_\Delta (x) < 0, f_\omega (x) < 0$] for $x > 0$ leads to
$T_{\rm c} > T_{{\rm c}0}$.
Once the two levels are split, the exchange scattering process gives rise to an attractive interaction
for pairing, by analogy with the optical phonon effects in superconductivity.
\cite{Fulde70}
\par

We would like to mention a linear combination of eqs.~(\ref{eqn:f0}) and (\ref{eqn:fs}),
[$f_0 (x) / \tau_{12}^0 + f_{\rm s} (x) / \tau_{12}^{\rm s}$],
which is derived from the singlet-multiplet configuration since both magnetic and nonmagnetic
scatterings coexist.
In this case, the weight of $f_\Delta (x)$ becomes smaller in the combined equation, so that $x$ has to
be sufficiently large to satisfy the condition $|f_\Delta (x)| > |f_\omega (x)|$ for \Tc~enhancement.
The details will be examined for the singlet-doublet configuration in \S3.2.
\par

\subsection{Extension to multiband superconductors}

\subsubsection{Bulk property}

Before extending the above single-band case to a two-band one,
we briefly review the earlier works by Shul {\it et al}.
\cite{Shul59}
and Kondo
\cite{Kondo63}
for multiband superconductivity.
The unique property of the multiband is described by the following model Hamiltonian:
\begin{align}
& \H_{\rm bulk} = \sum_{\mu = \pm} \sum_{\sigma = \uparrow, \downarrow} \sum_{\bsk}
\epsilon_{\bsk \mu} c_{\bsk \mu \sigma}^\dagger c_{\bsk \mu \sigma} \cr
&~~~~~~
+ \sum_{\mu,\mu'=\pm} \sum_{\bsk\bsk'} V_{\mu\mu'}
c_{\bsk \mu \uparrow}^\dagger c_{-\bsk \mu \downarrow}^\dagger
c_{-\bsk' \mu' \downarrow} c_{\bsk' \mu' \uparrow}.
\end{align}
Here, the first term represents the kinetic energy for the $\mu(=\pm)$ conduction band and
$c_{\bsk \mu \sigma}^\dagger$ ($c_{\bsk \mu \sigma}$) is the creation (annihilation) operator.
The second term represents the interaction between electrons with coupling constants $V_{\mu\mu'}$:
$\mu=\mu'$ for the intraband and $\mu\neq\mu'$ for the interband.
\Tc~is determined by the following form of the linearized gap equation:
\begin{align}
& \left(
  \begin{array}{cc}
    -V_{++} N_{0+} & -V_{+-} N_{0-} \cr
    -V_{-+} N_{0+} & -V_{--} N_{0-}
  \end{array}
\right)
\left(
  \begin{array}{c}
    \Delta_+ \cr
    \Delta_-
  \end{array}
\right) \cr
&~~~~~~
=\frac{1}{\log\left[ ( 2\re^\gamma\omega_{\rm c} ) / ( \pi T_{\rm c} ) \right]}
\left(
  \begin{array}{c}
    \Delta_+ \cr
    \Delta_-
  \end{array}
\right).
\label{eqn:gap-equation-Shul}
\end{align}
Here, $\Delta_\pm$ and $N_{0\pm}$ are the order parameters and density of states at the Fermi
energy for the $\mu=\pm$ band, respectively.
$\gamma(\simeq 0.577)$ is Euler's constant and we have assumed the same cut-off energy
$\omega_{\rm c}$ ($\gg T_{\rm c}$) for the two bands.
\Tc~corresponds to a positive eigenvalue of the $2\times 2$ matrix on the left-hand side of
eq.~(\ref{eqn:gap-equation-Shul}).
The multiband superconductivity is characterized by whether the interband interaction is attractive or
repulsive.
To see this point clearly, let us consider a simple case of two identical bands,
where $V_{++}=V_{--}<0$, $V_{+-}=V_{-+}$, and $N_{0+}=N_{0-}=N_0$.
The solution of eq.~(\ref{eqn:gap-equation-Shul}) is obtained as
\begin{align}
& T_{\rm cA} = \frac{2\re^\gamma\omega_{\rm c}}{\pi} \exp \left[ - \frac{1}{(|V_{++}| - V_{+-})N_{0}}
\right], \cr
& T_{\rm cB} = \frac{2\re^\gamma\omega_{\rm c}}{\pi} \exp \left[ - \frac{1}{(|V_{++}| + V_{+-})N_{0}} \right].
\label{eqn:Tc-AB}
\end{align}
Here, $T_{\rm cA}$ and $T_{\rm cB}$ are the transition temperatures for
\begin{align}
\Delta_{\rm A} = \frac{\Delta_+ + \Delta_-}{2},~~~~~~
\Delta_{\rm B} = \frac{\Delta_+ - \Delta_-}{2},
\label{eqn:Delta-AB}
\end{align}
respectively.
When $V_{+-}<0$, $T_{\rm cA}>T_{\rm cB}$ ($T_{\rm cB}$ = 0 if $|V_{+-}| > |V_{++}|$).
This means that the higher \Tc~is due to the attractive interband interaction for the order parameters with
the same sign ($\Delta_+ \Delta_- > 0$).
Conversely, when $V_{+-}>0$, the higher \Tc~is obtained for their opposite signs
($\Delta_+ \Delta_- < 0$).
The former and latter are called the $s_{++}$-wave and $s_\pm$-wave states, respectively.
\par

\subsubsection{Impurity effect}

Although electron-phonon and Coulomb interactions were taken into account as possible origins
of the interband interaction before,
\cite{Shul59,Kondo63}
we here propose inelastic (dynamical) impurity scattering as another origin that can give rise to
\Tc~enhancement in multiband superconductors even if it is due to magnetic impurities.
For this purpose, we extend the work by Fulde {\it et al}.
to a two-band case.
\par

Then, we reexamine the gap equation in eq.~(\ref{eqn:single-gap}) for a two-band $s$-wave
superconducting state with order parameters,
$\Delta_\mu$ for the $\mu$ band.
In the absence of impurities, the superconducting transition temperatures are expressed as $T_{{\rm c}0\pm}$
for the $\mu=\pm$ bands, respectively.
Here, the interband interaction is not taken into account for simplicity ($V_{+-}=0$).
Considering both magnetic and nonmagnetic scattering processes, we obtain the following linearized gap equation in the two-band case:
\begin{align}
& \frac{8 T_{\rm c}}{\pi}
\left(
  \begin{array}{c}
    \Delta_+ \log{\ds{\frac{T_{\rm c}}{T_{{\rm c}0+}}}} \cr
    \Delta_- \log{\ds{\frac{T_{\rm c}}{T_{{\rm c}0-}}}}
  \end{array}
\right) \cr
&~~
= \sum_{{\rm X} = 0, {\rm s}} \frac{1}{\tau_{12}^{\rm X}}
\left[ f_\Delta (x) \blambda_\Delta^{\rm X} + f_\omega (x) \blambda_\omega^{\rm X} \right]
\left(
  \begin{array}{c}
    \Delta_+ \cr
    \Delta_-
  \end{array}
\right),
\end{align}
where \Tc~is the transition temperature in the presence of impurities.
The energy difference $x$ between the impurity singlet ground and multiplet excited states is scaled
by $2T_{\rm c}$ [see eq.~(\ref{eqn:x})].
The matrices $\blambda_\Delta^{\rm X}$ and $\blambda_\omega^{\rm X}$ express
the impurity intraband and interband scattering contributions to the self-energies,
$\Sigma_\Delta (\ri \omega)$ and $\Sigma_\omega (\ri \omega)$, respectively:
\begin{align}
\blambda_\xi^{\rm X} =
\left(
  \begin{array}{cc}
    \lambda_{\xi,++}^{\rm X} & \lambda_{\xi,+-}^{\rm X} \cr
    \lambda_{\xi,-+}^{\rm X} & \lambda_{\xi,--}^{\rm X}
  \end{array}
\right)~~(\xi = \Delta, \omega).
\label{eqn:lambda}
\end{align}
In the following study, it is convenient to use the order parameters given in eq. (\ref{eqn:Delta-AB}),
where $\Delta_{\rm A}$ and $\Delta_{\rm B}$ are order parameters for the $s_{++}$-wave and
$s_\pm$-wave states, respectively.
By the unitary transformation
\begin{align}
\bLambda_\xi^{\rm X} = U^{-1} \blambda_\xi^{\rm X} U,~~~~~~
U = \frac{1}{\sqrt{2}}
\left(
\begin{array}{cc}
1 & 1 \\
1 & -1
\end{array}
\right),
\label{eqn:unitary}
\end{align}
the linearized gap equation is rewritten as
\begin{align}
& \left[ \frac{1}{2} \log\left(\frac{T_{{\rm c}0+}}{T_{{\rm c}0-}}\right)
\left(
\begin{array}{cc}
0 & 1 \\
1 & 0
\end{array}
\right)
 + \bLambda (x) \right] \left(
  \begin{array}{c}
    \Delta_{\rm A} \\
    \Delta_{\rm B}
  \end{array}
\right) \cr
&~~~~~~
= \frac{8 T_{\rm c}}{\pi} \log\left(\frac{T_{\rm c}}{\sqrt{T_{{\rm c}0+}T_{{\rm c}0-}}}\right)
\left(
  \begin{array}{c}
    \Delta_{\rm A} \\
    \Delta_{\rm B}
  \end{array}
\right),
\label{eqn:gap-eqn0} \\
&~~~~~~
\bLambda (x) = \sum_{{\rm X} = 0, {\rm s}} \frac{1}{\tau_{12}^{\rm X}}
\sum_{\xi = \Delta, \omega} f_\xi (x) \bLambda_\xi^{\rm X}.
\nonumber
\end{align}
We can see that $T_{\rm c}$, which is the superconducting transition temperature in the presence of impurities,
plays a role as an eigenvalue of the matrix on the left-hand side in eq. (\ref{eqn:gap-eqn0}).
The largest eigenvalue corresponds to the highest $T_{\rm c}$.
When $T_{{\rm c}0+}=T_{{\rm c}0-}$ for simplicity, only $\bLambda (x)$ is left on the left-hand side,
and a positive eigenvalue of $\bLambda$ leads to $T_{\rm c}$ enhancement.
\par

Next, we consider a case of two identical bands with interband interaction ($V_{+-}\neq 0$).
In this case, the \Tc~values are different for the $s_{++}$-wave and $s_\pm$-wave states, as given by
eq.~(\ref{eqn:Tc-AB}).
In the absence of impurities, we introduce transition temperatures $T_{{\rm c}0{\rm A}}$ and
$T_{{\rm c}0{\rm B}}$ for the $s_{++}$ and $s_\pm$-wave states, respectively.
For \Tc~in the presence of impurities, the gap equation is expressed as
\begin{align}
& \left[ \frac{1}{2} \log\left(\frac{T_{{\rm c}0{\rm A}}}{T_{{\rm c}0{\rm B}}}\right)
\left(
\begin{array}{cc}
1 & 0 \\
0 & -1
\end{array}
\right)
+ \bLambda (x) \right] \left(
  \begin{array}{c}
    \Delta_{\rm A} \\
    \Delta_{\rm B}
  \end{array}
\right) \cr
&~~~~~~
=\frac{8 T_{\rm c}}{\pi} \log\left(\frac{T_{\rm c}}{\sqrt{T_{{\rm c}0{\rm A}}T_{{\rm c}0{\rm B}}}}\right)
\left(
  \begin{array}{c}
    \Delta_{\rm A} \\
    \Delta_{\rm B}
  \end{array}
\right).
\label{eqn:gap-eqn}
\end{align}
Here, the contribution of the impurities is expressed by $\bLambda(x)$ and it modifies the
transition temperatures in the bulk.
\par

In the following part of this paper, we consider only two identical bands without the interband scattering
($V_{+-}=0$)
to capture the essence of the $T_{\rm c}$ enhancement caused by the inelastic scattering impurities,
where $T_{{\rm c}0{\rm A}}=T_{{\rm c}0{\rm B}}=T_{{\rm c}0}$.
We note that it is easy to extend the formulation to $T_{{\rm c}0{\rm A}}\neq T_{{\rm c}0{\rm B}}$ cases.
\par

\subsection{Example of magnetic interband scattering for \Tc~enhancement in $s_\pm$-wave state}

As mentioned above, \Tc~is enhanced by inelastic nonmagnetic impurity scattering in single-band
$s$-wave superconductors.
In this subsection, we show that magnetic interband scattering can also cause \Tc~enhancement in
multiband cases.
For this purpose, we focus on the roles of spin-dependent scattering in the $s_\pm$-wave state,
which is suggested as one of possible superconducting states realized in Fe pnictide superconductors.
\cite{Kuroki08,Mazin08}
As we will see, the interband scattering is important for \Tc~enhancement, which is unique to the
multiband and is never seen in the single-band case.
In fact, neither intraband nor elastic impurity scattering can be neglected in real systems.
These effects cause \Tc~suppression.
Whether \Tc~is enhanced or not depends on how to balance the pairing and depairing effects
caused by impurities.
\par

Here, we extend the formulation given in \S2.1 straightforwardly.
The conduction electron part for the $\mu$ band is given by
\begin{align}
& \H_{\mu} = \sum_\sigma \int\rd\br \psi_{\mu \sigma}^\dagger(\br) \epsilon(-\ri\nabla)
 \psi_{\mu\sigma}(\br)
\cr
&~~~~~~ - \Delta_\mu \int\rd\br \left[ \psi_{\mu \uparrow}^\dagger(\br)\psi_{\mu \downarrow}^\dagger(\br)
                                           + \psi_{\mu \downarrow}(\br)\psi_{\mu \uparrow}(\br) \right] \cr
&~~~~~~~~~~~~ (\mu=+,-).
\end{align}
Here, $\psi_{\mu \sigma}(\br)$ is a field operator of the conduction election for the $\mu$ band.
$\Delta_\mu$ is the $\mu$ band superconducting order parameter
for the $s_{\pm}$-wave superconductivity,  where $\Delta_\mu$ takes a real value.
For spin-dependent intraband and interband scatterings at the impurities, the interaction Hamiltonian
is defined by
\begin{align}
& \H' = \sum_{\bsR_\gamma} \sum_{mn} \sum_{\mu \nu} \sum_{\sigma\sigma'} \int\rd\br
     a_{\gamma m}^\dagger a_{\gamma n} \delta(\br-\bR_\gamma) \cr
&~~~~~~\times
      \bM_{mn,\mu\nu}\cdot\bbsigma_{\sigma\sigma'} \psi_{\mu \sigma}^\dagger(\br) \psi_{\nu \sigma'}(\br),
\label{eqn:H'}
\end{align}
where $\bbsigma$ is the Pauli matrix with the three components $\bsigma_\alpha$ ($\alpha = 1, 2, 3$).
As the corresponding spin exchange, $\bM_{mn,\mu\nu}$ is the scattering matrix element that
depends on the band components as well.
\par

We introduce the following $4\times 4$ matrix form of the thermal Green's function for the $\mu$ band:
\begin{align}
\bG_\mu (\tau,\br,\br') = - \langle T \bPsi_\mu (\br,\tau) \bPsi_\mu^\dagger(\br',0) \rangle,
\end{align}
where $\bPsi_\mu (\br)$ and $\bPsi_\mu^\dagger(\br)$ are defined for each band as
\begin{align}
& \bPsi_\mu (\br) =
   \left(
     \begin{array}{c}
       \psi_{\mu \uparrow}(\br) \\
       \psi_{\mu \downarrow}(\br) \\
       \psi_{\mu \uparrow}^\dagger(\br) \\
       \psi_{\mu \downarrow}^\dagger(\br)
     \end{array}
   \right), \cr
& \bPsi_\mu^\dagger(\br) =
   \left(
     \begin{array}{cccc}
       \psi_{\mu \uparrow}^\dagger(\br) &  \psi_{\mu \downarrow}^\dagger(\br) & \psi_{\mu \uparrow}(\br)
       & \psi_{\mu \downarrow}(\br)
     \end{array}
   \right).
\label{eqn:Psi}
\end{align}
Their Heisenberg representations $\bPsi_\mu (\br, \tau)$ and $\bPsi_\mu^\dagger (\br, \tau)$ are given
in the same manner as eq.~(\ref{eqn:Heisenberg}).
After introducing the unperturbed Green's function
\begin{align}
\bG_\mu (\ri\om,\bk)
  = - \frac{ \ri\om + \epsilon_{\bsk}\brho_3 + \Delta_\mu \brho_2 \bsigma_2   }
   { \omd + \epsilon_{\bsk}^2 + \Delta_\mu^2 },
\end{align}
as in eq.~(\ref{eqn:G88-0}), and
combining the two-band forms, we use the following $8\times 8$ matrix form of Green's function:
\begin{align}
\bG_0(\ri\om,\bk) =
\left(
  \begin{array}{cc}
    \bG_+(\ri\om,\bk) & 0 \cr
    0 & \bG_-(\ri\om,\bk)
  \end{array}
\right).
\end{align}
In particular, for the $s_\pm$-wave state ($\Delta_+ =- \Delta_-=\Delta$), it is rewritten as
\begin{align}
\bG_0(\ri\om,\bk) = - \frac{ \ri\om + \epsilon_\bsk \brho_3 + \Delta\btau_3\brho_2\bsigma_2 }{ \omd + \epsilon_\bsk^2 + \Delta^2 },
\label{eqn:G88}
\end{align}
where $\btau_\alpha$ ($\alpha=1,2,3$) is the Pauli matrix for the band space.
Similarly, the matrix for impurity scattering is given by, for instance,
\begin{align}
& \bU = \sum_{mn} \sum_{\alpha=x,y,z} \bU_{mn}^\alpha, \cr
&~~~~~~ \bU_{mn}^x = M_{mn}^x \btau_1\brho_3\bsigma_1,~~
\bU_{mn}^y = M_{mn}^y \btau_1\bsigma_2, \cr
&~~~~~~
\bU_{mn}^z = M_{mn}^z \btau_1\brho_3\bsigma_3.
\label{eqn:U-matrix}
\end{align}
Here, $\btau_1$ represents the interband scattering.
For the scattering matrix $\bU$ in eq.~(\ref{eqn:U-matrix}), the self-energy in Fig.~\ref{fig:selfenergy}
is given by
\begin{align}
& \bSigma(\ri\om) = - n_{\rm imp}T^2 \sum_{mn}
  \sum_{\omega_1\omega_2} \frac{1}{\ri\omega_1-\delta_m} \frac{1}{\ri\omega_2-\delta_n} \cr
&~~~~~~\times
  \frac{1}{\Omega} \sum_\bsk \sum_\alpha \bU_{mn}^\alpha \bG_0(\ri\om+\ri\omega_1-\ri\omega_2,\bk) \bU_{nm}^\alpha.
\end{align}
As in the single-band case, $\Delta$ in eq.~(\ref{eqn:G88}) changes its sign by the
$\bU_{mn}^\alpha \bG_0\bU_{nm}^\alpha$
transformation.
This is a key to the \Tc~enhancement by inelastic scattering also in the multiband case.
\par

\begin{figure}[t]
\begin{center}
\includegraphics[width=7cm]{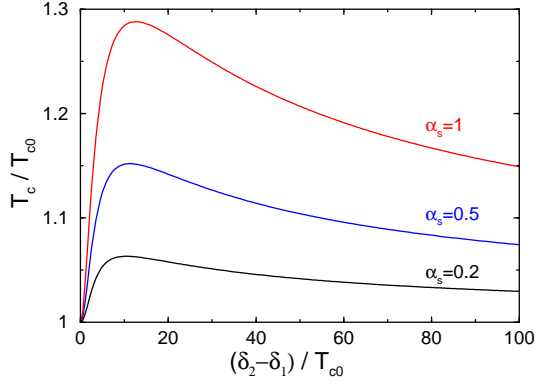}
\end{center}
\caption{
(Color online)
$(\delta_2-\delta_1)$ dependence of \Tc~for various values of $\alpha_{\rm s}$.
}
\label{fig:tc}
\end{figure}

We apply the above argument to the spin-dependent interband scattering case such as eq.~(\ref{eqn:U-matrix})
that enhances the \Tc~of the $s_\pm$-wave superconductivity.
In the calculation of the self-energy, the scattering matrix $\bU_{mn}^\alpha$ satisfies
\begin{align}
& \bU_{mn}^\alpha (\Delta \btau_3 \brho_2 \bsigma_2) \bU_{nm}^\alpha =
- | M_{mn}^\alpha |^2 (\Delta \btau_3 \brho_2 \bsigma_2), \cr
& \bU_{mn}^\alpha \bU_{nm}^\alpha = | M_{mn}^\alpha |^2,
\label{eqn:U-Delta}
\end{align}
which leads to
\begin{align}
& \bLambda_\Delta^{\rm s} =
\left(
\begin{array}{cc}
1 & 0 \\
0 & -1
\end{array}
\right),~~
\bLambda_\omega^{\rm s} =
\left(
\begin{array}{cc}
1 & 0 \\
0 & 1
\end{array}
\right), \\
& \frac{1}{\tau_{12}^{\rm s}} = 2\pi n_{\rm imp} N_0 \sum_\alpha \left| M_{12}^\alpha  \right|^2,
\end{align}
in eq. (\ref{eqn:gap-eqn}).
Here, $N_0$ represents the density of normal electron states at the Fermi energy.
We have assumed here a singlet-singlet configuration for the energy levels ($m,n = 1,2$) of the
impurity.
Then, we obtain the following gap equation:
\begin{align}
\alpha_{\rm s}
\left(
\begin{array}{cc}
- f_{\rm s} (x) & 0 \\
0 & f_0 (x)
\end{array}
\right) \left(
  \begin{array}{c}
    \Delta_{\rm A} \\
    \Delta_{\rm B}
  \end{array}
\right)
= \frac{T_{\rm c}}{T_{{\rm c}0}} \log\frac{T_{\rm c}}{T_{{\rm c}0}}
\left(
  \begin{array}{c}
    \Delta_{\rm A} \\
    \Delta_{\rm B}
  \end{array}
\right),
\end{align}
where $x=(\delta_2-\delta_1)/(2T_{\rm c})$ and $\alpha_{\rm s}$ represents the strength of the
spin-dependent impurity scattering defined by
\begin{align}
\alpha_{\rm s}=\frac{\pi}{8T_{\rm c0}\tau_{12}^{\rm s}}.
\end{align}
In the gap equation, $f_0 (x) = - f_\Delta (x) + f_\omega (x) > 0$ and $f_{\rm s} (x) = - f_\Delta (x) - f_\omega (x) > 0$
[see eqs.~(\ref{eqn:f0}) and (\ref{eqn:fs})].
This means that \Tc~is enhanced for the $s_\pm$-wave state.
We show the ($\delta_2-\delta_1$) dependence of \Tc~for various $\alpha_{\rm s}$ values
in Fig. \ref{fig:tc}.
At $\delta_1=\delta_2$, where the two impurity states are degenerate, there is no \Tc~enhancement.
When ($\delta_2-\delta_1$) is increased, \Tc~increases accordingly and takes a maximum value.
In fact, \Tc~depends on two factors competing with each other.
One is the strength of the attractive interaction between electrons as derived in the BCS theory.
The other is the energy region for the attractive interaction related to the cutoff.
The former is intensified by a small ($\delta_2-\delta_1$),
while the latter becomes large for a large ($\delta_2-\delta_1$).
In Fig. \ref{fig:tc}, one can see the maximum at approximately
($\delta_2-\delta_1)\simeq 10 T_{\rm c0}$.
For ($\delta_2-\delta_1)\rightarrow\infty$, there is no enhancement in \Tc,
since such a higher-lying energy level does not contribute to the attractive interaction.
\par

Thus, the order of $(\delta_2 - \delta_1) \sim T_{\rm c}$ is the most appropriate for \Tc~enhancement
by the interband scattering in the $s_{\pm}$-wave state.
On the other hand, it must be pointed out that \Tc~suppression is caused by elastic scattering due to
the impurity singlet ground state or by intraband scattering [for instance, $\btau_1$ is replaced by
unity in eq.~(\ref{eqn:U-matrix})], which have been neglected here.
The most important point is the internal structure of impurities that intensifies the interband magnetic
scattering to overcome these pair-breaking effects.
\par

\section{Typical Impurity Interband Scattering for \Tc~Enhancement}

In \S2.3, we have discussed \Tc~enhancement due to the spin-dependent interband scattering in
$s_\pm$-wave states as one of the examples, where the details of such impurity scattering have been
put aside.
In this section, we study typical impurity scattering for a singlet-multiplet configuration
that can give rise to \Tc~enhancement in multiband superconductors.
First, we show possible examples for the singlet-singlet and next apply the same argument to
the singlet-multiplet case.
In \S3.2, we take account of not only the interband impurity scattering effect but also the intraband
impurity scattering effect neglected in the previous section.
\par

\subsection{Singlet-singlet configuration}

In addition to the $\btau_1$ type in eq.~(\ref{eqn:U-matrix}) for the interband scattering, there is another
type of magnetic scattering, $M_{mn} \btau_2$, due to an orbital moment (spin-independent scattering).
Since it satisfies
\begin{align}
\btau_2 (\btau_3 \brho_2 \bsigma_2) \btau_2 = - (\btau_3 \brho_2 \bsigma_2),
\end{align}
\Tc~can be enhanced by the impurity scattering for the $s_\pm$ wave.
Here, we consider realistic cases for
$\bU_{mn}^z = M_{mn}^z \btau_1\brho_3\bsigma_3$ and $\bU_{mn} = M_{mn} \btau_2$.
In practice, we check the possible symmetry of $M_{12}$ for the singlet-singlet configuration
corresponding to each electron scattering type, assuming local orbital symmetries of band
electrons at the impurity site.
Although actual bands can include several orbital components, we represent each band by
an orbital component that mainly contributes to the band construction.
\par

First, let us begin with the spin-dependent case, $\bU_{mn}^z = M_{mn}^z \btau_1\brho_3\bsigma_3$.
Within the subspace,
$\bpsi_b = (\psi_{+\uparrow}~~\psi_{+\downarrow}~~\psi_{-\uparrow}~~\psi_{-\downarrow})^t$,
where $t$ denotes transposition, $\bU_{mn}^z$ is given as
\begin{align}
\bU_{mn}^z = M_{mn}^z \btau_1 \bsigma_3 = M_{mn}^z
\left(
\begin{array}{cccc}
 0 & 0 & 1 & 0 \\
 0 & 0 & 0 & -1 \\
 1 & 0 & 0 & 0 \\
 0 & -1 & 0 & 0
\end{array}
\right).
\label{eqn:umnz}
\end{align}
To identify the symmetry of electron scattering that depends on orbital components, we here
assume the $yz$ and $xz$ orbital types for the $\mu = +$ and $\mu = -$ bands, respectively.
The wave functions $\psi_{yz,\sigma}$ and $\psi_{xz,\sigma}$ at an impurity are connected to
$j = 3/2$ angular momentum bases $\psi_{j_z}$ (specifically, the $O_h$ $\Gamma_8$
point-group bases) by the following unitary transformation as
\begin{align}
& \left(
 \begin{array}{c}
 \psi_{yz \uparrow} \\
 \psi_{xz \uparrow} 
\end{array}
\right) = \frac{1} {\sqrt{2}}
\left(
\begin{array}{cc}
\ri & \ri \\
1 & -1
\end{array}
\right)
\left(
\begin{array}{c}
 \psi_{3/2} \\
 \psi_{-1/2}
\end{array}
\right), \cr
& \left(
 \begin{array}{c}
 \psi_{yz \downarrow} \\
 \psi_{xz \downarrow}
\end{array}
\right) = \frac{1}{\sqrt{2}}
\left(
\begin{array}{cc}
- \ri & - \ri \\
-1 & 1
\end{array}
\right)
\left(
\begin{array}{c}
 \psi_{1/2} \\
 \psi_{-3/2}
\end{array}
\right).
\end{align}
In this new basis set
$\bpsi_{j = 3/2} = (\psi_{3/2}~~\psi_{1/2}~~\psi_{-1/2}~~\psi_{-3/2})^t$,
it is useful to classify the scattering types (dipole, quadrupole, and octupole).
$\bU_{mn}^z$ in eq.~(\ref{eqn:umnz}) is transformed as
\begin{align}
\bU_{mn}^z~~\rightarrow~~M_{mn}^z
\left(
\begin{array}{cccc}
 0 & 0 & \ri & 0 \\
 0 & 0 & 0 & -\ri \\
 -\ri & 0 & 0 & 0 \\
 0 & \ri & 0 & 0
\end{array}
\right)_{j = 3/2~{\rm space}}.
\label{eqn:xyz-type}
\end{align}
This matrix expression corresponds to the $xyz$ type of tensor for $j = 3/2$.
In the local scattering at the impurity, $M_{mn}^z$ has the same symmetry as $xyz$.
In the cubic point group, $M_{mn}^z$ expresses $O_h$ $\Gamma_2$ octupole coupling
($\Gamma_3$ for $D_{4h}$ point group).
This coupling is realized in the $f^2$ configuration (doubly occupied $f$-electron state).
For instance, it connects the $D_{4h}$ crystal-field ground state
\begin{align}
| g \rangle = c_1 ( | 4 \rangle + | -4 \rangle ) + c_2 | 0 \rangle~~~~~~
( 2|c_1|^2 + |c_2|^2 = 1),
\label{eqn:state1}
\end{align}
with the first excited state
\begin{align}
| e \rangle = \frac{1}{\sqrt{2}} ( | 2 \rangle + | -2 \rangle ),
\end{align}
in the inelastic impurity scattering $\bU_{mn}^z$.
Here, $| M \rangle$ ($M = -4 \sim 4$) is an eigenstate of $J_z$ for the $J = 4$ angular momentum
state in the $f^2$ configuration.
Within the two crystal-field states, the scattering matrix is expressed by
\begin{align}
\left(
\begin{array}{cc}
 M_{11}^z & M_{12}^z \\
 M_{21}^z & M_{22}^z
\end{array}
\right) = |M_{12}^z|
\left(
\begin{array}{cc}
 0 & -\ri \\
 \ri & 0 
\end{array}
\right).
\end{align}
\par

Next, we consider the spin-independent case, $\bU_{mn} = M_{mn} \btau_2$.
In the subspace $\bpsi_b$, we obtain
\begin{align}
\bU_{mn} = M_{mn}
\left(
\begin{array}{cccc}
 0 & 0 & - \ri & 0 \\
 0 & 0 & 0 & - \ri \\
 \ri & 0 & 0 & 0 \\
 0 & \ri & 0 & 0
\end{array}
\right).
\label{eqn:Umn}
\end{align}
Assuming the $yz$ and $xz$ types for the two bands in this case as well, $\bU_{mn}$ is transformed to
the $j = 3/2$ angular momentum basis expression as
\begin{align}
\bU_{mn}~~\rightarrow~~M_{mn}
\left(
\begin{array}{cccc}
 -1 & 0 & 0 & 0 \\
 0 & -1 & 0 & 0 \\
 0 & 0 & 1 & 0 \\
 0 & 0 & 0 & 1
\end{array}
\right)_{j = 3/2~{\rm space}},
\end{align}
which expresses the combination of the $z$ dipole and $z(2z^2 - 3x^2 -3y^2)$ octupole types
of $j = 3/2$ electron scattering.
For the inelastic impurity scattering in this case, $M_{mn}$ represents the dipole coupling
between two low-lying singlet states, e.g., the ground state $| g \rangle$ in eq.~(\ref{eqn:state1}) and
the excited state,
\begin{align}
| e \rangle = \frac{1}{\sqrt{2}} ( | 4 \rangle - | - 4 \rangle ),
\end{align}
given as
\begin{align}
\left(
\begin{array}{cc}
 M_{11} & M_{12} \\
 M_{21} & M_{22}
\end{array}
\right) = |M_{12}|
\left(
\begin{array}{cc}
 0 & 1 \\
 1 & 0 
\end{array}
\right).
\end{align}
\par

Finally, we mention another pair of two-band types, ($2 z^2 - x^2 - y^2$) and ($x^2 - y^2$).
Applying the above argument, we find that the $\btau_1 \brho_3 \bsigma_3$ type of
electron scattering is reduced to the $z(x^2 - y^2)$ octupole type, and the $\btau_2$ type corresponds
to the $xyz$ octupole type expressed by eq.~(\ref{eqn:xyz-type}).
In Appendix~A, we give a different analysis for $\bU_{mn}^z$.
\par

\subsection{Singlet-doublet configuration}

Here, we devote ourselves to extending the present theory of \Tc~enhancement
to typical examples of impurities with internal degrees of freedom.
Orbital degrees of freedom of conduction electrons give rise to magnetic and nonmagnetic
exchange scatterings, both of which are considered here in the singlet-multiplet configuration.
Whether \Tc~is enhanced or suppressed depends on the ratio of their scattering strengths
that determines the signs of the two-band superconducting order parameters:
they are the same ($\Delta_+ \Delta_- > 0$) or different ($\Delta_+ \Delta_- < 0$).
For the \Tc~enhancement, the crystal-field ground state must be a singlet (a nonmagnetic doublet
is also allowed).
As mentioned for the $s_{\pm}$-wave state ($\Delta_+ \Delta_- < 0$) in \S2.3, \Tc~suppression is
caused by elastic scattering due to the singlet, which is neglected here as well.
First, we discuss a case of singlet-doublet configuration regarded as an $S = 1$ local pseudo-spin.
This $S = 1$ spin is not a spin triplet but a spin and orbitally coupled state, as often studied for
$f$-electron systems.
The latter can be realized as an impurity low-lying state in a uniaxial ($D_{4h}$ or $D_{6h}$)
crystal field.
In the same framework, we tackle a more complicated case of singlet-triplet configuration
discussed in the next subsection.
\par

For spin and orbitally coupled impurity states, in general, local orbital exchange occurs
as well as spin exchange during electron scattering by the impurity moment.
In the case of an $S^{\rm I} = 1$ pseudo-spin for an impurity, a spherical type of exchange interaction is expressed by coupling with, for instance, local $S^{\rm c} = 3/2$ states,
$\bpsi^{\rm c} = (\psi_{3/2}~~\psi_{1/2}~~\psi_{-1/2}~~\psi_{-3/2})^t$,
formed by conduction electrons.
This local interaction Hamiltonian is given by
\cite{Koga08}
\begin{align}
& H_{\rm loc} = \bpsi^{{\rm c} \dagger} \left( J_S \bS^{\rm I} \cdot \bS^{\rm c}
+ J_Q \sum_{\eta = 1}^5 Q_\eta^{\rm I} Q_\eta^{\rm c} \right) \bpsi^{\rm c} \cr
&~~~~~~+ \frac{1}{3} \Delta_{\rm CF} \left[ 3 \left( S_z^{\rm I} \right)^2 - 2 \right],
\label{eqn:H-loc}
\end{align}
where a potential (elastic) scattering term is neglected.
The first and second terms represent dipolar and quadrupolar exchanges with the coupling constants,
$J_S$ and $J_Q$, respectively.
The quadrupole operators ($Q_\eta^{\rm I}$ for an impurity; $Q_\eta^{\rm c}$ for an electron)
are defined as
\begin{align}
& \{Q_\eta, \eta = 1, \cdots , 5 \} \cr
&~~~~~~ =\{S_y S_z + S_z S_y, S_z S_x + S_x S_z, S_x S_y + S_y S_x, \cr
&~~~~~~~~~~~~ S_x^2 - S_y^2, (2 S_z^2 - S_x^2 - S_y^2) / \sqrt{3} \}.
\end{align}
The last term in eq.~(\ref{eqn:H-loc}) introduces uniaxial ($D_{4h}$ or $D_{6h}$) anisotropy to the
impurity states, and $\Delta_{\rm CF}$ ($>0$) is taken to determine a singlet ground state here.
For the $S^{\rm I} = 1$ pseudo-spin, it is sufficient to consider the above dipoles and quadrupoles.
We do not consider the anisotropy of each exchange coupling that usually exists in a realistic system,
which does not affect the following argument.
\par

On the other hand, we introduce some assumptions to electron states as follows.
At impurity sites, partial waves of conduction electrons are represented by $S^{\rm c} = 3/2$.
This can be regarded as the $O_h$ $\Gamma_8$ point-group basis
in a cubic system.
In general, actual conduction bands can include all four components of $S^{\rm c} = 3/2$, and
their mixing is expressed as
\begin{align}
\bpsi^{\rm c} = \bV \bpsi_b,~~~~~~
\bpsi^{\rm c} \equiv
   \left(
     \begin{array}{c}
       \psi_{3/2} \\
       \psi_{1/2} \\
       \psi_{-1/2} \\
       \psi_{-3/2}
     \end{array}
   \right),~~~~~~
\bpsi_b \equiv
   \left(
     \begin{array}{c}
       \psi_{+ \uparrow} \\
       \psi_{+ \downarrow} \\
       \psi_{- \uparrow} \\
       \psi_{- \downarrow}
     \end{array}
   \right),
\label{eqn:V-band1}
\end{align}
where each element in the transformation matrix $\bV$ is given by the overlap of local orbital
and band wave functions such as $\langle S_z^{\rm c} | \mu \sigma \rangle$
($\mu = \pm; \sigma = \uparrow, \downarrow$).
To examine the orbital roles in the superconductivity, we consider here the simplest case in which two
($\pm 3/2$) of the orbital components enter the $+$ band and the other two ($\pm 1/2$) enter
the $-$ band.
In the above $\bV$, we assume that
$ \langle 3/2 | + \downarrow \rangle = \langle 1/2 | - \uparrow \rangle = \langle -1/2 | - \downarrow \rangle
= \langle -3/2 | + \uparrow \rangle = V_0$ and that the other matrix elements vanish:
\begin{align}
\bV = V_0
\left(
\begin{array}{cccc}
0 & 1 & 0 & 0 \\
0 & 0 & 1 & 0 \\
0 & 0 & 0 & 1 \\
1 & 0 & 0 & 0
\end{array}
\right).
\label{eqn:V-band2}
\end{align}
Such a one to one correspondence clarifies the connection between the $S^{\rm c} = 3/2$
pseudo-spin space and the SU(2) spin $\otimes$ SU(2) band space, and the scattering matrices
are expressed simply by eqs.~(\ref{eqn:Smatrix}) and (\ref{eqn:Qmatrix}).
Although this simplification overestimates interband scattering compared with intraband scattering,
it helps us examine what types of interband correlations are relevant to the relative signs of
order parameters and how intraband scattering modifies their relative amplitudes.
\par

For the superconducting order parameters, we consider both the $s_{++}$ wave ($\Delta_+ = \Delta_-$)
and the $s_\pm$ wave ($\Delta_+ = - \Delta_-$).
Since we neglect any correlations between the two bands except for the impurity scattering,
the local correlations directly affect the relative signs of order parameters together with their
amplitudes.
This is justified when the impurity effect is more relevant than any other interband correlation such as
an interband Coulomb interaction.
We respectively express the $s_{++}$-wave and $s_\pm$-wave states as
\begin{align}
& \bDelta_{\rm A} = \Delta_{\rm A} \brho_2 \bsigma_2~~(s_{++}~{\rm wave}),
\label{eqn:Delta-A} \\
& \bDelta_{\rm B} = \Delta_{\rm B} \btau_3 \brho_2 \bsigma_2~~(s_\pm~{\rm wave}),
\label{eqn:Delta-B}
\end{align}
with the order parameters $\Delta_{\rm A}$ and $\Delta_{\rm B}$, respectively.
The matrix for the order parameter is given by their combination as
\begin{align}
\bDelta = \bDelta_{\rm A} + \bDelta_{\rm B}.
\label{eqn:Delta}
\end{align}
\par

The relevant impurity scatterings for the singlet-doublet configuration are described in Appendix~B.
First, we examine properties of the dipolar and quadrupolar scatterings separately.
Let us start from the dipolar (magnetic) scattering case where $J_S \ne 0$ and $J_Q = 0$ in
eq.~(\ref{eqn:H-loc}).
Applying $\H'$ in eq.~(\ref{eqn:Hex}) to calculate the self-energy in Fig.~\ref{fig:selfenergy}, we obtain
\begin{align}
\bSigma (\ri\om) = & - n_{\rm imp}T^2 \sum_{n \ne n'} \sum_{\omega_1\omega_2}
  \frac{1}{\ri\omega_1-\delta_n} \frac{1}{\ri\omega_2-\delta_{n'}} \left( \sqrt{2} \right)^2 \cr
& \times \left( \frac{J_S}{2} \right)^2 \frac{1}{\Omega} \sum_\bsk
  \left( S_+^{\rm c} \bG_0 S_-^{\rm c} + S_-^{\rm c} \bG_0 S_+^{\rm c} \right),
\end{align}
where $\bG_0 \equiv \bG_0 (\ri\om+\ri\omega_1-\ri\omega_2,\bk)$.
$\delta_1$ ($\delta_2$) corresponds to the energy level of the impurity singlet ground
(doublet excited) state ($n,n' = 1,2$) and $\delta_2 - \delta_1 = \Delta_{\rm CF}$.
The factor $(\sqrt{2})^2$ comes from the second-order process $S_+ S_-$ or $S_- S_+$ between
the singlet and doublet states.
Then, we derive a gap equation.
As in eq.~(\ref{eqn:G88}), the Green's function $\bG_0$ consists of three parts:
$\ri\om$, $\epsilon_\bsk\brho_3$, and $\bDelta$.
The $\epsilon_\bsk\brho_3$ term disappears after the summation over $\bk$.
The other two parts are transformed by
($S_+^{\rm c} \bG_0 S_-^{\rm c} + S_-^{\rm c} \bG_0 S_+^{\rm c}$).
For $\bDelta$,
\begin{align}
& S_+^{\rm c} \bDelta S_-^{\rm c} + S_-^{\rm c} \bDelta S_+^{\rm c} \cr
&~~~~~~ = (5\Delta_{\rm A} - 2\Delta_{\rm B})\brho_2 \sigma_2 + (- 2\Delta_{\rm A} - \Delta_{\rm B})
\btau_3 \brho_2 \bsigma_2 \cr
&~~~~~~ = \Delta_{\rm A'} \brho_2 \sigma_2 + \Delta_{\rm B'} \btau_3 \brho_2 \bsigma_2.
\end{align}
This means that the order parameters are transformed as
\begin{align}
\left(
\begin{array}{c}
\Delta'_{\rm A} \\
\Delta'_{\rm B}
\end{array}
\right)
= \bLambda_\Delta^S
\left(
\begin{array}{c}
\Delta_{\rm A} \\
\Delta_{\rm B}
\end{array}
\right),~~~~~~
\bLambda_\Delta^S =
\left(
\begin{array}{cc}
5 & -2 \\
-2 & -1
\end{array}
\right).
\end{align}
Similarly, the $\ri\om$ part is
\begin{align}
S_+^{\rm c} \ri\om S_-^{\rm c} + S_-^{\rm c} \ri\om S_+^{\rm c}
= \ri\om ( 5 - 2 \btau_3 ).
\end{align}
Since $\btau_3$ is the Pauli matrix for the $\pm$ band space,
eq.~(\ref{eqn:lambda}) is obtained as
\begin{align}
\blambda_\omega^S =
\left(
\begin{array}{cc}
3 & 0 \\
0 & 7
\end{array}
\right),
\end{align}
for $X = S$ and $\xi = \omega$.
Then, eq.~(\ref{eqn:unitary}) yields
\begin{align}
\bLambda_\omega^S = \bU^{-1} \blambda_\omega^S \bU =
\left(
\begin{array}{cc}
5 & -2 \\
-2 & 5
\end{array}
\right).
\end{align}
Both $\bLambda_\Delta^S$ and $\bLambda_\omega^S$ can be divided into interband and intraband scattering parts as follows:
\begin{align}
& \bLambda_\Delta^S =
\left(
\begin{array}{cc}
3 & 0 \\
0 & -3
\end{array}
\right) +
\left(
\begin{array}{cc}
2 & -2 \\
-2 & 2
\end{array}
\right), \cr
& \bLambda_\omega^S =
\left(
\begin{array}{cc}
3 & 0 \\
0 & 3
\end{array}
\right) +
\left(
\begin{array}{cc}
2 & -2 \\
-2 & 2
\end{array}
\right).
\end{align}
If the intraband contribution, the second matrix on the right-hand side of each equation, is removed,
the interband contribution results in the \Tc~enhancement for the $s_\pm$ state, as discussed in \S2.3.
\par

In the same manner, the above argument is applied to the quadrupolar (nonmagnetic) scattering case
where $J_S = 0$ and $J_Q \ne 0$ in eq.~(\ref{eqn:H-loc}).
For the $\bDelta$ and $\ri \om$ parts in the self-energy,
\begin{align}
& Q_+^{\rm c} \bDelta Q_-^{\rm c} + Q_-^{\rm c} \bDelta Q_+^{\rm c} =
- 12 \Delta_{\rm A} \brho_2 \sigma_2 + 12 \Delta_{\rm B} \btau_3 \brho_2 \bsigma_2, \\
& Q_+^{\rm c} \ri\om Q_-^{\rm c} + Q_-^{\rm c} \ri\om Q_+^{\rm c}
= 12 \ri\om,
\end{align}
respectively, and we obtain
\begin{align}
\bLambda_\Delta^Q =
\left(
\begin{array}{cc}
-12 & 0 \\
  0 & 12
\end{array}
\right),~~
\bLambda_\omega^Q =
\left(
\begin{array}{cc}
12 & 0 \\
 0 & 12
\end{array}
\right).
\end{align}
These lead to the \Tc~enhancement for the $s_{++}$-wave state.
\par

\begin{figure}[t]
\begin{center}
\includegraphics[width=7cm,clip]{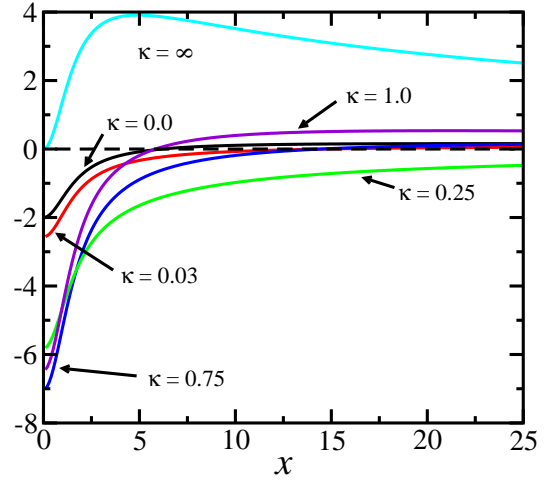}
\end{center}
\caption{
(Color online)
$x$ [$= \Delta_{\rm CF} / (2T_{\rm c})$] dependence of the highest eigenvalue of the matrix
$\bar{\tau}_{12} \bLambda (x)$ for \Tc.
The plot is shown for fixed $\kappa = J_Q^2 / J_S^2$ values.
}
\label{fig:eigenvalue}
\end{figure}

\begin{figure}[t]
\begin{center}
\includegraphics[width=6cm,clip]{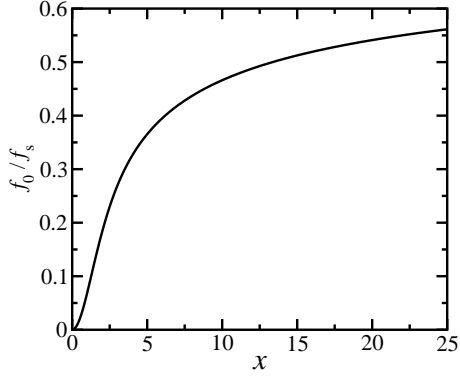}
\end{center}
\caption{
Ratio $f_0 / f_{\rm s}$ as a function of $x = \Delta_{\rm CF} / (2T_{\rm c})$.
}
\label{fig:f0fs}
\end{figure}

\begin{figure}[t]
\begin{center}
\includegraphics[width=6cm,clip]{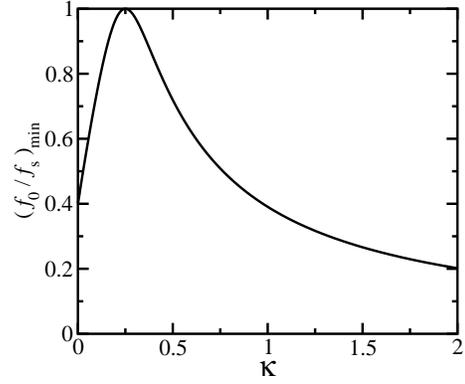}
\end{center}
\caption{
$\kappa$ ($= J_Q^2 / J_S^2$) dependence of the minimum of the ratio $f_0 / f_{\rm s}$
for $T_{\rm c}$ enhancement.
}
\label{fig:f0fsmin}
\end{figure}

Next, we see the more generic treatment combining both magnetic ($S$) and nonmagnetic
($Q$) scattering terms as
\begin{align}
& \bar{\tau}_{12} \bLambda (x) =
\cos \zeta \left[ f_\Delta (x) \bLambda_\Delta^S
+ f_\omega (x) \bLambda_\omega^S
\right] \cr
&~~~~~~~~~~~~~~~~~~ + \sin \zeta \left[ f_\Delta (x) \bLambda_\Delta^Q
+ f_\omega (x) \bLambda_\omega^Q
\right] \cr
&~~
= \left(
\begin{array}{cc}
[ - 5 f_{\rm s} (x) \cos \zeta & 2 f_{\rm s} (x) \cos \zeta \\
~~~~~~ + 12 f_0 (x) \sin \zeta ] & \\
& \{ [ 3 f_0 (x) - 2 f_{\rm s} (x) ] \cos \zeta \\
2 f_{\rm s} (x) \cos \zeta & -12 f_{\rm s} (x) \sin \zeta \}
\end{array}
\right), \cr
&
\end{align}
in eq.~(\ref{eqn:gap-eqn}), where
\begin{align}
\bar{\tau}_{12} = \frac{\tau_{12}^S \tau_{12}^Q}
{\sqrt{\Big( \tau_{12}^S \Big)^2 + \Big( \tau_{12}^Q \Big)^2}},~~
\tan \zeta = \frac{J_Q^2}{J_S^2} \equiv \kappa,
\end{align}
and the lifetime $\tau_{12}^S$ ($\tau_{12}^Q$) is introduced for the magnetic (nonmagnetic)
scattering.
The highest eigenvalue of $\bar{\tau}_{12} \bLambda$ determines the quantity of
\begin{align}
\frac{8 T_{\rm c}}{\pi} \bar{\tau}_{12} \log \frac{T_{\rm c}}{T_{{\rm c}0}},
\label{eqn:Tc}
\end{align}
and its $x$ dependence is shown in Fig.~\ref{fig:eigenvalue}.
The \Tc~enhancement is obtained for its positive value, which holds at $x > 6.2$ for $\kappa = 0$
and at $x > 5.8$ for $\kappa = 1.0$.
In the presence of only nonmagnetic scattering for $\kappa = \infty$, \Tc~is enhanced in the
entire $x > 0$ range.
The \Tc~suppression, found for a small $x$ or $\kappa \simeq 0.25$, is due to the competition
between the magnetic and nonmagnetic exchange scatterings and to the depairing effect by the
intraband scattering.
Since $f_0 / f_{\rm s} < 1$, the condition for a positive value of
eq.~(\ref{eqn:Tc}) is given by
\begin{align}
( - 5 f_{\rm s} + 12 \kappa f_0 ) ( 3 f_0 - 2 f_{\rm s} - 12 \kappa f_{\rm s} ) -  ( 2 f_{\rm s} )^2 < 0,
\end{align}
the solution of which determines the minimum $f_0 / f_{\rm s}$ as a function of $\kappa$:
\begin{align}
& \left( \frac{f_0}{f_{\rm s}} \right)_{\rm min} = \frac{1}{24 \kappa}
\left[ ( 48 \kappa^2 + 8 \kappa + 5 ) \right. \cr
&~~~~~~\left.
- \sqrt{ ( 48 \kappa^2 + 8 \kappa + 5 )^2
- 96 \kappa ( 10 \kappa + 1 ) } \right].
\label{eqn:f0fsmin}
\end{align}
The $x$ dependence of $f_0 / f_{\rm s}$ and the $\kappa$ dependence of $(f_0 / f_{\rm s})_{\rm min}$
are shown in Figs.~\ref{fig:f0fs} and \ref{fig:f0fsmin}, respectively.
At both $\kappa = 0.03$ and $\kappa = 0.75$, for instance, eq.~(\ref{eqn:f0fsmin}) takes almost
the same value, $\simeq 0.51$, which gives the minimum $x \simeq 14.5$ for \Tc~enhancement, as
shown in Fig.~\ref{fig:eigenvalue}.
The maximum at $\kappa = 0.25$ in Fig.~\ref{fig:f0fsmin} indicates that there is no \Tc~enhancement
for any finite crystal-field level splitting, implying that \Tc~is always suppressed
by the competition between the magnetic and nonmagnetic scattering effects.
We also find that $\kappa$ determines the relative signs of the two order parameters:
$\Delta_+ \Delta_- < 0$ for $\kappa < 0.25$ and $\Delta_+ \Delta_- > 0$ for $\kappa > 0.25$.
The combination of $\bDelta_{\rm A}$ ($s_{++}$ wave) and $\bDelta_{\rm B}$ ($s_\pm$ wave) is
caused by the intraband scattering effect in the $-$ band, due to the off-diagonal elements in
$\bLambda (x)$, which leads to $| \Delta_- | < | \Delta_+ |$ in the vicinity of $T_{\rm c}$.
At $\kappa = 0.25$, $\Delta_- = 0$ ($\Delta_{\rm A} = \Delta_{\rm B}$) means that only one band
($+$ band) is superconducting with \Tc~suppression.
\par

The above argument is based on a rather artificial assumption about the local band character in
eq.~(\ref{eqn:V-band2}).
More generic treatment of $\hat{V}$ in eq.~(\ref{eqn:V-band1}) provides us with various scattering
effects on \Tc.
This point is considered for the singlet-triplet configuration discussed below.
\par

\subsection{Singlet-triplet configuration}

The singlet-triplet configuration is realized for a non-Kramers ion in an $O_h$ crystal field environment
like the Pr$^{3+}$ or U$^{4+}$ $f^2$ low-lying states in heavy-fermion materials.
For the strong spin-orbit coupling, the most relevant local $f$-electron states are described by the
$j = 5/2$ angular momentum.
Then, we consider only the exchange coupling between the impurity states and the $j = 5/2$ electrons
hybridized with conduction bands.
We assume here that the $\Gamma_8$ and $\Gamma_7$ partial waves (see Appendix~C) are
transferred independently to the $+$ and $-$ bands, respectively.
In terms of eq.~(\ref{eqn:notation}) for both $\Gamma_8$ and $\Gamma_7$,
$\bpsi_m = (\psi_{m_1 \uparrow}~~\psi_{m_1 \downarrow}~~\psi_{m_2 \uparrow}
~~\psi_{m_2 \downarrow}~~\psi_{m_3 \uparrow}~~\psi_{m_3 \downarrow})^t$
is combined with
$\bpsi_b = (\psi_{+\uparrow}~~\psi_{+\downarrow}~~\psi_{-\uparrow}~~\psi_{-\downarrow})^t$
for the bands.
This is expressed by $\bpsi_m = \bV \bpsi_b$:
\begin{align}
\bV =
\left(
\begin{array}{cccc}
v_+ u_{1 \uparrow} & v_+ v_{1 \uparrow} & 0 & 0 \\
v_+ v_{1 \downarrow} & v_+ u_{1 \downarrow} & 0 & 0 \\
v_+ u_{2 \uparrow} & v_+ v_{2 \uparrow} & 0 & 0 \\
v_+ v_{2 \downarrow} & v_+ u_{2 \downarrow} & 0 & 0 \\
0 & 0 & v_- & 0 \\
0 & 0 & 0 & v_-
\end{array}
\right).
\label{eqn:m87-b+-}
\end{align}
Here, $v_\pm$ represents the hybridization amplitude of the $f$-orbitals and the $\mu = \pm$ band,
respectively, at the impurity sites;
\begin{align}
& u_{i \uparrow} = \langle m_i \uparrow | + \uparrow \rangle,~~
v_{i \uparrow} = \langle m_i \uparrow | + \downarrow \rangle, \cr
& v_{i \downarrow} = \langle m_i \downarrow | + \uparrow \rangle,~~
u_{i \downarrow} = \langle m_i \downarrow | + \downarrow \rangle
\end{align}
represent the impurity site overlaps of the wave functions $\psi_{m_i, \sigma}$ ($i = 1,2$) and
$\psi_{+, \sigma}$.
We only simplify the connection between the $m_3$ orbital and the $-$ band.
\par

In Appendix~C, the relevant impurity scatterings are described for the singlet-triplet configuration
$\Gamma_1 \oplus \Gamma$ ($\Gamma = \Gamma_4, \Gamma_5$).
For $\H'$ in eq.~(\ref{eqn:Hex2}), the self-energy in Fig.~\ref{fig:selfenergy} is obtained as
\begin{align}
\bSigma^\Gamma (\ri\om) = & - n_{\rm imp}T^2 \sum_{n \ne n'} \sum_{\omega_1\omega_2}
  \frac{1}{\ri\omega_1-\delta_n} \frac{1}{\ri\omega_2-\delta_{n'}} \cr
& \times \sum_{X = S,Q}  \left( J_X^\Gamma \right)^2 \frac{1}{\Omega} \sum_\bsk
  \left\{ X_z^{\rm c, \Gamma}  \bG_0 X_z^{\rm c, \Gamma} \right. \cr
& \left.
  + \frac{1}{2} \left[ X_+^{\rm c, \Gamma} \bG_0 X_-^{\rm c, \Gamma}
  + X_-^{\rm c, \Gamma} \bG_0 X_+^{\rm c, \Gamma} \right] \right\} \cr
& [ \bG_0 \equiv \bG_0 (\ri\om+\ri\omega_1-\ri\omega_2,\bk) ].
\end{align}
Here, $\delta_1$ ($\delta_2$) corresponds to the energy level of the impurity singlet ground
(triplet excited) state ($n,n' = 1,2$).
For the gap equation in eq.~(\ref{eqn:gap-eqn}), the transformation to $\bDelta$ in
eq.~(\ref{eqn:Delta}),
\begin{align}
X_z^{{\rm c}, \Gamma} \bDelta X_z^{{\rm c}, \Gamma}
+ \frac{1}{2} ( X_+^{{\rm c}, \Gamma} \bDelta X_-^{{\rm c}, \Gamma}
+ X_-^{{\rm c}, \Gamma} \bDelta X_+^{{\rm c}, \Gamma} ),
\end{align}
leads to
\begin{align}
\bLambda_\Delta^S = (-1) \bLambda_\Delta^Q
= \frac{3}{8} \sum_{i = 1}^2 ( u_{i \uparrow} u_{i \downarrow} - v_{i \uparrow} v_{i \downarrow} )
\left(
\begin{array}{cc}
-1 & 0 \\
0 & 1
\end{array}
\right),
\label{eqn:LambdaD-SQ}
\end{align}
in both $\Gamma = \Gamma_4$ and $\Gamma = \Gamma_5$ cases.
For the derivation, we use
$u_{i \uparrow} u_{i \downarrow} = u_{i \uparrow}^* u_{i \downarrow}^*$ and
$v_{i \uparrow} v_{i \downarrow} = v_{i \uparrow}^* v_{i \downarrow}^*$ since the $m_{i \uparrow}$
and $m_{i \downarrow}$ local electrons are the time reversal partners.
In the same manner, for both $\Gamma_4$ and $\Gamma_5$,
\begin{align}
X_z^{{\rm c}, \Gamma} \ri \om X_z^{{\rm c}, \Gamma}
+ \frac{1}{2} ( X_+^{{\rm c}, \Gamma} \ri \om X_-^{{\rm c}, \Gamma}
+ X_-^{{\rm c}, \Gamma} \ri \om X_+^{{\rm c}, \Gamma} )
\end{align}
gives
\begin{align}
\bLambda_\omega^S = \bLambda_\omega^Q
= \frac{3}{16} \sum_{i=1}^2 \sum_{\sigma = \uparrow, \downarrow}
( u_{i \sigma}^* u_{i \sigma} + v_{i \sigma}^* v_{i \sigma} )
\left(
\begin{array}{cc}
1 & 0 \\
0 & 1
\end{array}
\right).
\label{eqn:Lambdao-SQ}
\end{align}
Equation~(\ref{eqn:LambdaD-SQ}) implies the competition between the magnetic ($S$) and
nonmagnetic ($Q$) scattering effects for \Tc~enhancement.
Which has the higher \Tc~, the $s_{++}$ wave or the $s_\pm$ wave, depends on whether the sign of
($u_{i \uparrow} u_{i \downarrow} - v_{i \uparrow} v_{i \downarrow}$) is positive or negative,
respectively.
In realistic systems, both magnetic and nonmagnetic scatterings coexist.
In the present case, we can calculate the coupling constants, $J_S^\Gamma$ and $J_Q^\Gamma$,
based on the Anderson model including the $j = 5/2$ electron exchange scattering due to a single
impurity in the $f^2$ configuration,
\cite{Koga95}
and obtain the ratio as $| J_Q^\Gamma | / | J_S^\Gamma | = 1 / 3$.
\cite{Koga06}
This means that the magnetic interband scattering dominates the \Tc~enhancement for the
singlet-triplet configuration.
A more detailed discussion is given below.
\par

\section{Pr Impurity Effect in \La~Superconductor}

The Pr$^{3+}$ $f^2$ configuration is a good candidate for raising \Tc~if Pr can be embedded in a
multiband superconductor.
In fact, for the skutterudite superconductor \LaPr, the Pr singlet-triplet configuration may be relevant to \Tc~enhancement in \La.
\cite{Chang07}
Here, we show an attempt to apply the above argument in this case.
The most intriguing feature of the skutterudite compounds is that each rare-earth ion is located at the
center of the pnictogen cage (Sb$_{12}$) having the $a_u$($xyz$) and $t_u$($x,y,z$) molecular
orbitals.
It is considered that the Pr $f$-electron states hybridize with the conduction bands via these orbitals.
For a strong spin-orbit coupling, the $a_u$ electrons have the $O_h$ $\Gamma_7$ symmetry
and transfer directly to the $m_3$ electron state named in eq.~(\ref{eqn:notation}).
On the other hand, the $t_u$ electrons with the $O_h$ $\Gamma_8$ symmetry mix with both $m_1$
and $m_2$ states as
\cite{Onodera66}
\begin{align}
& | m_1, \uparrow \rangle \leftrightarrow \frac{1}{\sqrt{2}}
( | x, \downarrow \rangle - \ri | y, \downarrow \rangle ), \\
& | m_1, \downarrow \rangle \leftrightarrow - \frac{1}{\sqrt{2}}
( | x, \uparrow \rangle + \ri | y, \uparrow \rangle ), \\
& | m_2, \uparrow \rangle \leftrightarrow \frac{1}{\sqrt{3}}
   \left[ \sqrt{2} | z, \uparrow \rangle - \frac{1}{\sqrt{2}}
   ( | x, \downarrow \rangle + \ri | y, \downarrow \rangle ) \right], \\
& | m_2, \downarrow \rangle \leftrightarrow \frac{1}{\sqrt{3}}
   \left[ \sqrt{2} | z, \downarrow \rangle + \frac{1}{\sqrt{2}}
   ( | x, \uparrow \rangle - \ri | y, \uparrow \rangle ) \right].
\end{align}
We here consider two conduction bands, one of which is $a_u$-dominant and the other is
$t_u$-dominant.
We assume that both are combined with each other only through interband electron scattering.
In the present case, the most relevant is the Pr impurity scattering due to the hybridization effect with
the bands.
Then, we can use the transformation in eq.~(\ref{eqn:m87-b+-}) for mixing $f$-electron states
with the $t_u$-dominant band ($+$ band) and the $a_u$-dominant band ($-$ band).
Choosing the local $t_u$($x,y,z$) component that mainly contributes to the $+$ band and
taking their onsite overlaps arbitrarily as
\begin{align}
\langle x | + \rangle : \langle y | + \rangle : \langle z | + \rangle
= \sin \theta \cos \phi : \sin \theta \sin \phi : \cos \theta,
\end{align}
we have
\begin{align}
& u_{1 \uparrow} = u_{1 \downarrow} = 0, \cr
& v_{1 \uparrow} = ( 1 / \sqrt{2} ) \sin \theta~ {\rm e}^{\ri \phi},~~
   v_{1 \downarrow} = - ( 1 / \sqrt{2} ) \sin \theta~ {\rm e}^{- \ri \phi}, \cr
& u_{2 \uparrow} = u_{2 \downarrow} = ( \sqrt{2 / 3} ) \cos \theta, \cr
&  v_{2 \uparrow} = - ( 1 / \sqrt{6} ) \sin \theta~ {\rm e}^{-\ri \phi},~~
    v_{2 \downarrow} = ( 1 / \sqrt{6} ) \sin \theta~ {\rm e}^{\ri \phi}.
\end{align}
Substituting them in eqs.~(\ref{eqn:LambdaD-SQ}) and (\ref{eqn:Lambdao-SQ}), we obtain
\begin{align}
& \bLambda_\Delta^S = (-1) \bLambda_\Delta^Q = \frac{1}{4}
\left(
\begin{array}{cc}
-1 & 0 \\
0 & 1
\end{array}
\right), \\
& \bLambda_\omega^S = \bLambda_\omega^Q = \frac{1}{4}
\left(
\begin{array}{cc}
1 & 0 \\
0 & 1
\end{array}
\right).
\end{align}
In the gap equation (\ref{eqn:gap-eqn}), we have
\begin{align}
\bLambda (x) & = \frac{1}{4 \tau_{12}^S}
\left(
\begin{array}{cc}
f_0 (x) & 0 \\
0 & - f_{\rm s} (x)
\end{array}
\right) \cr
& + \frac{1}{4 \tau_{12}^Q}
\left(
\begin{array}{cc}
- f_{\rm s} (x) & 0 \\
0 & f_0 (x)
\end{array}
\right).
\label{eqn:Lambda-SQ}
\end{align}
In the real system, we must consider the $T_h$ symmetry that combines the Pr $O_h$ $\Gamma_4$
and $\Gamma_5$ triplet states as
\cite{Koga06,Takegahara01,Kuramoto09}
\begin{align}
| \Gamma_4^{(2)} \eta \rangle = \sqrt{ 1 - d^2 } | \Gamma_5 \eta \rangle
+ d | \Gamma_4 \eta \rangle~~(\eta = +,0,-),
\label{eqn:Th}
\end{align}
where $d$ ($0 < d < 1 / \sqrt{2}$) represents the deviation from the $O_h$ symmetry.
In the above calculation of the self-energy, we usually have additional terms, namely,
\begin{align}
& \left[ X_z^{{\rm c}, \Gamma_4} \bDelta X_z^{{\rm c}, \Gamma_5}
+ \frac{1}{2} ( X_+^{{\rm c}, \Gamma_4} \bDelta X_-^{{\rm c}, \Gamma_5}
+ X_-^{{\rm c}, \Gamma_4} \bDelta X_+^{{\rm c}, \Gamma_5} ) \right] \cr
&~~~~~~ + (\Gamma_4 \leftrightarrow \Gamma_5) ~~~~~~ (X^{\rm c} = s, q),
\end{align}
which vanish in this case.
They also vanish when $\bDelta$ is replaced with $\ri \omega$.
If the $a_u$ band hybridizes with the $f$-orbitals more strongly than the $t_u$ band, which means
$v_- \gg v_+$ here, the impurity interband scattering enhances \Tc~for a finite crystal-field splitting
$x$ [$= (\delta_2 - \delta_1) / (2T_{\rm c})$], satisfying either $f_0 / f_{\rm s} > \tau_{12}^S / \tau_{12}^Q$
or $f_0 / f_{\rm s} > \tau_{12}^Q / \tau_{12}^S$.
For the former, the \Tc~enhancement is possible in the $s_{++}$-wave state, and a higher \Tc~can
be realized as the magnetic scattering becomes more dominant.
In \S3.3, we have mentioned that $|J_Q| / |J_S| = 1/3$ ($\tau_{12}^S / \tau_{12}^Q = 1/9$) is satisfied
for $O_h$.
It also holds for $T_h$,
\cite{Koga06}
so that the gap equation chooses the $s_{++}$ wave for the \Tc~enhancement.
One may think that the quadrupolar scattering is the most relevant if $d$ is small in eq.~(\ref{eqn:Th}),
since the $\Gamma_1$-$\Gamma_5$ interchange dominates the exchange scattering.
For the $\Gamma_5$ scattering type, however, the octupolar scattering cannot be neglected in the
multiorbital exchange owing to the hybridization of $f$-electrons with the conduction bands.
Thus, the magnetic scattering can contribute to the \Tc~enhancement for the
$\Gamma_1$-$\Gamma_4^{(2)}$ configuration in a multiband system.
Our result indicates that the $s_{++}$-wave state is favorable for the \LaPr~superconductivity
if the multiband picture is applicable and the Pr or Sb$_{12}$ site local orbital symmetries reflect
in the bands.
\par

Finally, we mention the effect of intraband scattering neglected in the above argument where
$v_- \gg v_+$ is assumed for the hybridization amplitudes.
If the $m_1$-$m_2$ scattering terms in eqs.~(\ref{eqn:qz}) and (\ref{eqn:q+}) are considered in
calculating the self-energy for the $\Gamma_5$ nonmagnetic scattering type, a correction term,
\begin{align}
C_v \frac{1}{\tau_{12}^Q} f_0 (x)
\left(
\begin{array}{cc}
1 & 1 \\
1 & 1
\end{array}
\right)~~~~~~\left( C_v \sim \frac{v_+^2}{v_-^2} \right),
\end{align}
is added on the right-hand side in eq.~(\ref{eqn:Lambda-SQ}).
It assists the \Tc~enhancement, which resembles the effect of inelastic nonmagnetic scattering
impurities in single-band $s$-wave superconductors.

\section{Conclusion}
We have studied inelastic (dynamical) impurity scattering effects on \Tc~enhancement in the two-band
superconducting states, $s_{++}$ wave and $s_\pm$ wave.
The key is to check the sign change of the order parameters in the self-energy corresponding to
the second order of the scattering process.
We solve a gap equation, where both the $s_{++}$ wave and $s_\pm$ wave are combined by
impurity scattering, and find out the possible atomic structure of the impurity that increases \Tc.
For the \Tc~enhancement, the crystal-field ground state must be a singlet or a nonmagnetic doublet.
If not, magnetic impurities will always cause \Tc~suppression.
For the $s_{\pm}$ wave, it is necessary for elastic (nonmagnetic) scattering to be relatively small.
In the singlet-singlet configuration, we can easily determine what type of magnetic interband scattering
contributes to pairing interaction.
This simple analysis is very useful for determining a scattering type for the \Tc~enhancement among
multiorbital interaction terms when we consider  such a complicated atomic structure as
the $f^n$ configuration.
Here, we have discussed the singlet-doublet and singlet-triplet configurations.
We find that the \Tc~of either the $s_{++}$-wave or $s_\pm$-wave state can be enhanced by
the dynamical magnetic or nonmagnetic impurity interband scattering for a larger crystal-field
splitting of impurity ground and excited states, while \Tc~is suppressed if both scattering strengths
are comparable.
Whether \Tc~is enhanced or not also depends on how the local electron states hybridize with the
two bands.
We show a case of \Tc~enhancement in the $s_{++}$ wave by magnetic interband
scattering due to the singlet-triplet configuration. 
This result may give useful information on a multiband picture of the \LaPr~superconductivity.
In fact, the connection between the La-rich and Pr-rich superconductors is left to be clarified
as well as the symmetry of the latter order parameter.
\cite{Aoki07}
\par

In the above argument, we have assumed weak scattering impurities.
Strictly speaking, dynamical scattering effects should be investigated as a Kondo problem.
It was shown theoretically in normal metallic cases that the impurity exchange scattering strength
is renormalized to be weaker with the decrease in temperature since the crystal-field singlet
competes with the Kondo-singlet formation.
\cite{Koga08,Koga96,Yotsuhashi02,Koga07}
This holds even for a small crystal-field splitting of the singlet ground and excited multiplet states.
Thus, our results shown here are valid for the \Tc~enhancement due to the impurity scattering.
\par

Throughout the paper, we focus on the hybridization between local orbitals and conduction
electron states.
For most of the $f$-electron systems, this may be more relevant than the admixture of orbitals
formed by intersite electron hopping that we have neglected here.
Our treatment is more practical since the latter contribution can be included effectively in the hybridization.
\par

The possibility of the $s_\pm$-wave state has been the subject of debate for
high-\Tc~superconductors
with FeAs layers.
It is pointed out that the Fe $d$-orbitals contribute to the disconnected Fermi surfaces and
the orbital degrees of freedom play an important role in pairing interaction.
\cite{Ishida09}
The interband Coulomb interactions may be relevant for the high \Tc~of this multiband
superconductivity,
\cite{Kondo63,Yamaji87,Yanagi09}
the roles of which are analogous to those of the local correlations in the $f$-electron systems
we have considered here.

\acknowledgement

This work is supported by a Grant-in-Aid for Scientific Research (No. 20540353) from the Japan Society for the Promotion of Science.
One of the authors (H. K.) is supported by a Grant-in-Aid for Scientific Research on Innovative Areas
``Heavy Electrons'' (No. 20102008)
from The Ministry of Education, Culture, Sports, Science and Technology, Japan.

\appendix
\section{Connection between the $s_\pm$-Wave and Conventional $s$-Wave
Superconductors}

We attempt to apply the unitary transformation
\begin{align}
\left(
 \begin{array}{c}
 \psi_{+, \sigma} \\
 \psi_{-, \sigma}
\end{array}
\right) = \frac{1}{\sqrt{2}}
\left(
\begin{array}{cc}
 1 & 1 \\
 -1 & 1
\end{array}
\right)
\left(
\begin{array}{c}
 \psi_{M \sigma} \\
 \psi_{-M \sigma}
\end{array}
\right)
\label{eqn:pmM}
\end{align}
to the band basis, where $M$ is a natural number.
Using this new basis set,
$\bpsi_M = (\psi_{M\uparrow}~~\psi_{M\downarrow}~~\psi_{-M\uparrow}~~\psi_{-M\downarrow})^t$,
$\bU_{mn}^z$ in eq.~(\ref{eqn:umnz}) is transformed as
\begin{align}
\bU_{mn}^z~~\rightarrow~~ - M_{mn}^z
\left(
\begin{array}{cccc}
 1 & 0 & 0 & 0 \\
 0 & -1 & 0 & 0 \\
 0 & 0 & -1 & 0 \\
 0 & 0 & 0 & 1
\end{array}
\right)_{M~{\rm space}}.
\label{eqn:umnz-quad}
\end{align}
If the number $M$ is regarded as an orbital component, this matrix indicates a quadrupole type of
scattering.
In fact, $\bpsi_{M = 1}$ corresponds directly to $\bpsi_{j = 3/2}$ in \S3.1.
In this case, the matrix in eq.~(\ref{eqn:umnz-quad}) is expressed by $(2 j_z^2 - j_x^2- j_y^2)$.
\par

On the other hand, for the conventional single-band $s$-wave superconductivity, the
pairing Hamiltonian is given by
\begin{align}
\H_\Delta = - \sum_\bsk \Delta
                     \left( c_{\bsk\uparrow}^\dagger c_{-\bsk\downarrow}^\dagger
                          + c_{-\bsk\downarrow} c_{\bsk\uparrow} \right).
\label{eqn:H-Delta}
\end{align}
After applying spherical expansion to the operators as
\begin{align}
c_{\bsk\sigma} = \sum_{LM} \ri^{-L} \frac{(6\pi)^{1/2}}{kR} Y_{LM}(\Omega_\bsk)c_{k L M\sigma},
\end{align}
where $R$ is the radius of the system, we rewrite the Hamiltonian (\ref{eqn:H-Delta}) as
\begin{align}
& \H_\Delta = - \sum_\sk \sum_{L, M \ge 0} (-1)^M \Delta \cr
&~~~~~~\times
                     \left( c_{\sk LM\uparrow}^\dagger c_{\sk L,-M\downarrow}^\dagger
                     + c_{\sk L,-M\downarrow} c_{\sk L M\uparrow} \right).
\label{eqn:H-D-LM}
\end{align}
This orbital component $M$ can correspond to $M$ in eq.~(\ref{eqn:pmM}) directly.
The pairing interaction works between the electrons with the different orbitals denoted
by ${\pm M}$.
Let us fix $L$ and restrict the orbitals to a pair of $\pm M$.
If $\psi_{M \sigma}$ is transformed to $\psi_{\mu \sigma}$ using eq.~(\ref{eqn:pmM}),
eq.~(\ref{eqn:H-D-LM}) is reduced to
\begin{align}
\H_\Delta = - \sum_{\mu = \pm} \sum_k (-1)^M \mu \Delta
                     \left( c_{k\mu\uparrow}^\dagger c_{k\mu\downarrow}^\dagger
                          + c_{k\mu\downarrow} c_{k\mu\uparrow} \right).
\end{align}
It represents $s_{\pm}$-wave pairing in the two orbitals connected to the $\mu$ bands.
\par

Thus the spin-dependent interband scattering in eq.~(\ref{eqn:umnz}) in a two-band $s_{\pm}$-wave
superconductor can be mapped to the quadrupole type of scattering in eq.~(\ref{eqn:umnz-quad})
in a single-band $s$-wave superconductor.
Both cases give rise to the \Tc~enhancement if the electron scattering interchanges the impurity
low-lying states with a finite crystal-field splitting.
\par

\section{Exchange Matrices for Singlet-Doublet Configuration}

At low temperatures, the most relevant terms in eq.~(\ref{eqn:H-loc}) are electron scatterings with
interchange between the local ground and excited states given by
\begin{align}
J_S \left[ S_x^{\rm I} S_x^{\rm c} + S_y^{\rm I} S_y^{\rm c} \right]
+ J_Q \left[ Q_1^{\rm I} Q_1^{\rm c} + Q_2^{\rm I} Q_2^{\rm c} \right].
\end{align}
Using $S_{\pm} = S_x \pm \ri S_y$ and $Q_{\pm} = \pm \ri Q_1 + Q_2$, it is rewritten as
\begin{align}
\frac{J_S}{2} \left[ S_+^{\rm I} S_-^{\rm c} + S_-^{\rm I} S_+^{\rm c} \right]
+ \frac{J_Q}{2} \left[ Q_+^{\rm I} Q_-^{\rm c} + Q_-^{\rm I} Q_+^{\rm c} \right].
\end{align}
By denoting $S_z^{\rm I} = 0$ (the impurity ground state) and $S_z^{\rm I} = 1, -1$
(the impurity excited states) by $| 1 \rangle$, $| 2 \rangle$, and $| 3 \rangle$, respectively,
the $(S_\pm^{\rm I})_{mn}$ and $(Q_\pm^{\rm I})_{mn}$ ($m,n = 1, 2, 3$) matrix expressions are
given as
\begin{align}
& S_+^{\rm I} = S_-^{{\rm I} \dagger} =
\left(
\begin{array}{ccc}
0 & 0 & \sqrt{2} \\
\sqrt{2} & 0 & 0 \\
0 & 0 & 0
\end{array}
\right),
\label{eqn:S-I} \\
& Q_+^{\rm I} = Q_-^{{\rm I} \dagger} =
\left(
\begin{array}{ccc}
0 & 0 & - \sqrt{2} \\
\sqrt{2} & 0 & 0 \\
0 & 0 & 0
\end{array}
\right).
\label{eqn:Q-I}
\end{align}
For the $S^{\rm c} = 3/2$ states, the operators $S_+^{\rm c} = S_-^{{\rm c} \dagger}$ and
$Q_+^{\rm c} = Q_-^{{\rm c} \dagger}$ are expressed by
\begin{align}
& \bpsi^{{\rm c} \dagger} S_+^{\rm c} \bpsi^{\rm c} = \bpsi^{{\rm c} \dagger}
\left(
\begin{array}{cccc}
0 & \sqrt{3} & 0 & 0 \\
0 & 0 & 2 & 0 \\
0 & 0 & 0 & \sqrt{3} \\
0 & 0 & 0 & 0
\end{array}
\right) \bpsi^{\rm c},
\label{eqn:S+} \\
& \bpsi^{{\rm c} \dagger} Q_+^{\rm c} \bpsi^{\rm c} = \bpsi^{{\rm c} \dagger}
\left(
\begin{array}{cccc}
0 & 2 \sqrt{3} & 0 & 0 \\
0 & 0 & 0 & 0 \\
0 & 0 & 0 & - 2 \sqrt{3} \\
0 & 0 & 0 & 0
\end{array}
\right) \bpsi^{\rm c}.
\end{align}
Following eq.~(\ref{eqn:H'}), we examine the low-temperature physics using the impurity interaction
Hamiltonian
\begin{align}
& \H' = \sum_{X = S,Q} \sum_{\bsR_\gamma} \sum_{mn}
         \int\rd\br a_{\gamma m}^\dagger a_{\gamma n} \delta(\br-\bR_\gamma) \cr
&~~~~~~\times
         \bPsi^\dagger(\br) \frac{J_X}{2} \left( X_{+,mn}^{\rm I} X_-^{\rm c}
         + X_{-, mn}^{\rm I} X_+^{\rm c} \right) \bPsi(\br). \cr
&
\label{eqn:Hex}
\end{align}
Here, $\bPsi$ is an eight-dimensional vector for conduction electrons that is obtained by extending
$\bpsi_b$ in eq.~(\ref{eqn:V-band1}) to the particle-hole space as
\begin{align}
\bPsi =
\left(
\begin{array}{c}
\bPsi_+ \\
\bPsi_-
\end{array}
\right),~~~~~~
\bPsi^\dagger =
\left(
\begin{array}{cc}
\bPsi_+^\dagger & \bPsi_-^\dagger
\end{array}
\right),
\label{eqn:Psi8}
\end{align}
where $\bPsi_\mu$ ($\mu = +, -$) is defined in eq.~(\ref{eqn:Psi}).
Through $\bV$ in eq.~(\ref{eqn:V-band2}) in this extended space
(Pauli matrices $\bsigma$ for spin, $\brho$ for particle hole, and $\btau$ for band),
the above electron scattering matrices are rewritten as
\begin{align}
& S_+^{\rm c} = \frac{\sqrt{3}}{2} \btau_1 ( \brho_3 \bsigma_1 - \ri \bsigma_2 )
+ \frac{1}{2} ( 1 - \btau_3 )( \brho_3 \bsigma_1 + \ri \bsigma_2 ),
\label{eqn:Smatrix} \\
& Q_+^{\rm c} = \sqrt{3} (\ri \btau_2)( \bsigma_1 - \ri \brho_3 \bsigma_2 ),
\label{eqn:Qmatrix}
\end{align}
where $|V_0|^2$ has been omitted.
One can see that $S_\pm^{\rm c}$ consists of the interband scattering $\tau_1$ and intraband scattering
$(1 - \btau_3)$ terms.
The former is comparable to $(\bU^x - \ri \bU^y)$ in eq.~(\ref{eqn:U-matrix}).
\par

\section{Exchange Matrices for Singlet-Triplet Configuration}

On the basis of the $J = 4$ total angular momentum, the $O_h$ singlet and triplet states are given by
\cite{Kuramoto09}
\begin{align}
& | \Gamma_1 \rangle = \frac{\sqrt{30}}{12} (| 4 \rangle + | -4 \rangle ) + \frac{\sqrt{21}}{6} | 0 \rangle, \\
& \left\{
\begin{array}{l}
| \Gamma_4 + \rangle = - \sqrt{\ds{\frac{1}{8}}} | -3 \rangle - \sqrt{\ds{\frac{7}{8}}} | 1 \rangle, \\
| \Gamma_4 0 \rangle = \sqrt{\ds{\frac{1}{2}}} (| 4 \rangle - | -4 \rangle ), \\
| \Gamma_4 - \rangle = \sqrt{\ds{\frac{1}{8}}} | 3 \rangle + \sqrt{\ds{\frac{7}{8}}} | -1 \rangle,
\end{array}
\right. \\
& \left\{
\begin{array}{l}
| \Gamma_5 + \rangle = \sqrt{\ds{\frac{7}{8}}} | 3 \rangle - \sqrt{\ds{\frac{1}{8}}} | -1 \rangle, \\
| \Gamma_5 0 \rangle = \sqrt{\ds{\frac{1}{2}}} (| 2 \rangle - | -2 \rangle ), \\
| \Gamma_5 - \rangle = -\sqrt{\ds{\frac{7}{8}}} | -3 \rangle + \sqrt{\ds{\frac{1}{8}}} | 1 \rangle.
\end{array}
\right.
\end{align}
For the $j = 5/2$ local electrons, they are classified into the $O_h$ symmetric states as
\cite{Onodera66}
\begin{align}
& \left\{
\begin{array}{l}
| \Gamma_{8,3/2} \rangle = - \sqrt{\ds{\frac{1}{6}}} | 3/2 \rangle - \sqrt{\ds{\frac{5}{6}}} | -5/2 \rangle, \\
| \Gamma_{8,1/2} \rangle = | 1/2 \rangle, \\
| \Gamma_{8,-1/2} \rangle = - | -1/2 \rangle, \\
| \Gamma_{8,-3/2} \rangle = \sqrt{\ds{\frac{1}{6}}} | -3/2 \rangle + \sqrt{\ds{\frac{5}{6}}} | 5/2 \rangle,
\end{array}
\right. \\
& \left\{
\begin{array}{l}
| \Gamma_{7,1/2} \rangle = \sqrt{\ds{\frac{5}{6}}} | -3/2 \rangle - \sqrt{\ds{\frac{1}{6}}} | 5/2 \rangle, \\
| \Gamma_{7,-1/2} \rangle = \sqrt{\ds{\frac{5}{6}}} | 3/2 \rangle - \sqrt{\ds{\frac{1}{6}}} | -5/2 \rangle. \\
\end{array}
\right.
\end{align}
In the following argument, we change their notations as
\begin{align}
& \Gamma_{8,-3/2} \rightarrow (m_1, \uparrow),~~
\Gamma_{8,3/2} \rightarrow (m_1, \downarrow),\cr
& \Gamma_{8,1/2} \rightarrow (m_2, \uparrow),~~
\Gamma_{8,-1/2} \rightarrow (m_2, \downarrow); \cr
& \Gamma_{7,1/2} \rightarrow (m_3, \uparrow),~~
\Gamma_{7,-1/2} \rightarrow (m_3, \downarrow).
\label{eqn:notation}
\end{align}
In the same manner as that in eq.~(\ref{eqn:S-I}) for an impurity, the magnetic coupling between
the $\Gamma_1$ singlet ground (denoted by $| 1 \rangle$) and
$\Gamma$ ($= \Gamma_4, \Gamma_5$) triplet
($+$, $0$ and $-$ states denoted by $| 2 \rangle$, $| 3 \rangle$ and $| 4 \rangle$, respectively) excited
states is expressed by the following matrices:
\cite{Koga06}
\begin{align}
& S_z^\Gamma =
\left(
\begin{array}{cccc}
0 & 0 & 1 & 0 \\
0 & 0 & 0 & 0 \\
1 & 0 & 0 & 0 \\
0 & 0 & 0 & 0
\end{array}
\right), \\
& S_+^\Gamma = S_-^{\Gamma \dagger} =
\left(
\begin{array}{cccc}
0 & 0 & 0 & \sqrt{2} \\
- \sqrt{2} & 0 & 0 & 0 \\
0 & 0 & 0 & 0 \\
0 & 0 & 0 & 0
\end{array}
\right).
\end{align}
For nonmagnetic coupling,
\begin{align}
& Q_z^\Gamma =
\left(
\begin{array}{cccc}
0 & 0 & - \ri & 0 \\
0 & 0 & 0 & 0 \\
\ri & 0 & 0 & 0 \\
0 & 0 & 0 & 0
\end{array}
\right), \\
& Q_+^\Gamma = Q_-^{\Gamma \dagger} =
\left(
\begin{array}{cccc}
0 & 0 & 0 & - \sqrt{2} \\
- \sqrt{2} & 0 & 0 & 0 \\
0 & 0 & 0 & 0 \\
0 & 0 & 0 & 0
\end{array}
\right).
\end{align}
We note that the relative signs of the off-diagonal matrix elements are different between
$S_\pm^\Gamma$ ($Q_\pm^\Gamma$) for the pseudo-quartet and $S_\pm^{\rm I}$ ($Q_\pm^{\rm I}$)
for the $S^{\rm I} = 1$ pseudo-spin given in eq.~(\ref{eqn:S-I}) [in eq.~(\ref{eqn:Q-I})].
The singlet-triplet interchange processes are mediated by the corresponding $\Gamma_8$ and
$\Gamma_7$ electron scatterings as follows:
\cite{Koga06}
\begin{align}
& \bpsi_m^\dagger s_z^{\Gamma_4} \bpsi_m =
\frac{1}{2} ( \psi_{m_1 \uparrow}^\dagger \psi_{m_3 \uparrow}
+ \psi_{m_1 \downarrow}^\dagger \psi_{m_3 \downarrow} ) + {\rm H. c.}, \\
& \bpsi_m^\dagger s_+^{\Gamma_4} \bpsi_m =
\frac{1}{2} ( - \psi_{m_1 \uparrow}^\dagger \psi_{m_3 \downarrow}
+ \psi_{m_3 \uparrow}^\dagger \psi_{m_1 \downarrow} ) \cr
&~~~~~~~~~~~~~~~~~~ + \frac{\sqrt{3}}{2} ( \psi_{m_2 \downarrow}^\dagger \psi_{m_3 \uparrow}
-  \psi_{m_3 \downarrow}^\dagger \psi_{m_2 \uparrow} ), \\
& \bpsi_m^\dagger s_z^{\Gamma_5} \bpsi_m =
\frac{1}{2} ( \psi_{m_2 \uparrow}^\dagger \psi_{m_3 \uparrow}
+ \psi_{m_2 \downarrow}^\dagger \psi_{m_3 \downarrow} ) + {\rm H. c.}, \\
& \bpsi_m^\dagger s_+^{\Gamma_5} \bpsi_m =
\frac{1}{2} ( - \psi_{m_2 \uparrow}^\dagger \psi_{m_3 \downarrow}
+ \psi_{m_3 \uparrow}^\dagger \psi_{m_2 \downarrow} ) \cr
&~~~~~~~~~~~~~~~~~~ + \frac{\sqrt{3}}{2} ( - \psi_{m_1 \downarrow}^\dagger \psi_{m_3 \uparrow}
+  \psi_{m_3 \downarrow}^\dagger \psi_{m_1 \uparrow} ), \\
& \bpsi_m^\dagger q_z^{\Gamma_4} \bpsi_m =
- \ri \frac{1}{2} ( \psi_{m_1 \uparrow}^\dagger \psi_{m_3 \uparrow}
+ \psi_{m_1 \downarrow}^\dagger \psi_{m_3 \downarrow} ) + {\rm H. c.}, \cr
& \\
& \bpsi_m^\dagger q_+^{\Gamma_4} \bpsi_m =
\frac{1}{2} ( \psi_{m_1 \uparrow}^\dagger \psi_{m_3 \downarrow}
+ \psi_{m_3 \uparrow}^\dagger \psi_{m_1 \downarrow} ) \cr
&~~~~~~~~~~~~~~~~~~ + \frac{\sqrt{3}}{2} ( - \psi_{m_2 \downarrow}^\dagger \psi_{m_3 \uparrow}
-  \psi_{m_3 \downarrow}^\dagger \psi_{m_2 \uparrow} ), \\
& \bpsi_m^\dagger q_z^{\Gamma_5} \bpsi_m =
- \ri \left\{ \frac{1}{2} ( \psi_{m_2 \uparrow}^\dagger \psi_{m_3 \uparrow}
+ \psi_{m_2 \downarrow}^\dagger \psi_{m_3 \downarrow} ) \right. \cr
&~~~~~~~~~~~~~~~~~~\left. + 2\sqrt{5} ( \psi_{m_1 \uparrow}^\dagger \psi_{m_2 \uparrow}
- \psi_{m_1 \downarrow}^\dagger \psi_{m_2 \downarrow} ) \right\} + {\rm H. c.},
\label{eqn:qz} \\
& \bpsi_m^\dagger q_+^{\Gamma_5} \bpsi_m =
\frac{1}{2} ( \psi_{m_2 \uparrow}^\dagger \psi_{m_3 \downarrow}
+ \psi_{m_3 \uparrow}^\dagger \psi_{m_2 \downarrow} ) \cr
&~~~~~~~~~~~~~~~~~~ + \frac{\sqrt{3}}{2} ( \psi_{m_1 \downarrow}^\dagger \psi_{m_3 \uparrow}
+  \psi_{m_3 \downarrow}^\dagger \psi_{m_1 \uparrow} ) \cr
&~~~~~~~~~~~~~~~~~~ + 4\sqrt{5} ( \psi_{m_1 \uparrow}^\dagger \psi_{m_2 \downarrow}
- \psi_{m_2 \uparrow}^\dagger \psi_{m_1 \downarrow} ).
\label{eqn:q+}
\end{align}
In \S3.3, we assume $v_- \gg v_+$ for the hybridization with the conduction band
in eq.~(\ref{eqn:m87-b+-}), i.e., we neglect the $m_1$-$m_2$ orbital scattering terms in
eqs.~(\ref{eqn:qz}) and (\ref{eqn:q+}), whose contribution is small in the band scattering.
Using the transformation in eq.~(\ref{eqn:m87-b+-}), the above multiorbital scattering operators are
reduced to interband scattering operators.
For the $\Gamma_4$ type,
\begin{align}
& \bpsi_b^\dagger s_z^{\Gamma_4} \bpsi_b =
\bpsi_b^\dagger \frac{1}{2}
\left(
\begin{array}{cccc}
0 & 0 & u_{1 \uparrow}^* & v_{1 \downarrow}^* \\
0 & 0 & v_{1 \uparrow}^* & u_{1 \downarrow}^* \\
u_{1 \uparrow} & v_{1 \uparrow} & 0 & 0 \\
v_{1 \downarrow} & u_{1 \downarrow} & 0 & 0
\end{array}
\right) \bpsi_b,
\label{eqn:s4z} \\
& \bpsi_b^\dagger s_+^{\Gamma_4} \bpsi_b =
{\small \bpsi_b^\dagger \frac{1}{2}
\left(
\begin{array}{cccc}
0 & 0 & \sqrt{3} v_{2 \downarrow}^* & - u_{1 \uparrow}^* \\
0 & 0 & \sqrt{3} u_{2 \downarrow}^* & - v_{1 \uparrow}^* \\
v_{1 \downarrow} & u_{1 \downarrow} & 0 & 0 \\
- \sqrt{3} u_{2 \uparrow} & - \sqrt{3} v_{2 \uparrow} & 0 & 0
\end{array}
\right) \bpsi_b }, \\
& \bpsi_b^\dagger q_z^{\Gamma_4} \bpsi_b =
\bpsi_b^\dagger \left( - \ri \frac{1}{2} \right)
\left(
\begin{array}{cccc}
0 & 0 & u_{1 \uparrow}^* & v_{1 \downarrow}^* \\
0 & 0 & v_{1 \uparrow}^* & u_{1 \downarrow}^* \\
- u_{1 \uparrow} & - v_{1 \uparrow} & 0 & 0 \\
- v_{1 \downarrow} & - u_{1 \downarrow} & 0 & 0
\end{array}
\right) \bpsi_b,
\label{eqn:q4z} \\
& \bpsi_b^\dagger q_+^{\Gamma_4} \bpsi_b =
{\small \bpsi_b^\dagger \frac{1}{2}
\left(
\begin{array}{cccc}
0 & 0 & - \sqrt{3} v_{2 \downarrow}^* & u_{1 \uparrow}^* \\
0 & 0 & - \sqrt{3} u_{2 \downarrow}^* & v_{1 \uparrow}^* \\
v_{1 \downarrow} & u_{1 \downarrow} & 0 & 0 \\
- \sqrt{3} u_{2 \uparrow} & - \sqrt{3} v_{2 \uparrow} & 0 & 0
\end{array}
\right) \bpsi_b },
\label{eqn:q4+}
\end{align}
where $(v_+ v_-)$, which appears as the common factor in each matrix element, has been omitted.
For the $\Gamma_5$ type, $s_z^{\Gamma_5}$ is obtained by switching the indices as
$1 \leftrightarrow 2$ in the $s_z^{\Gamma_4}$ matrix, and $s_+^{\Gamma_5}$ by replacing
$\sqrt{3} \rightarrow - \sqrt{3}$ in addition to switching $1\leftrightarrow 2$ in the
$s_+^{\Gamma_4}$ matrix;
$q_z^{\Gamma_5}$ and $q_+^{\Gamma_5}$ are obtained from $q_z^{\Gamma_4}$ and
$q_+^{\Gamma_4}$, respectively, in the same manner.
We note that $s_z^{\Gamma_4}$ in eq.~(\ref{eqn:s4z}) is comparable to $\bU_{mn}^z$ in
eq.~(\ref{eqn:umnz})
if $u_{1 \uparrow} = - u_{1 \downarrow} = u_{1 \uparrow}^* = - u_{1 \downarrow}^*$ and
$v_{1 \sigma} = 0$ are taken;
$q_z^{\Gamma_4}$ in eq.~(\ref{eqn:q4z}) is comparable to $\bU_{mn}$ in eq.~(\ref{eqn:Umn})
for $u_{1 \uparrow} = u_{1 \downarrow} = u_{1 \uparrow}^* = u_{1 \downarrow}^*$ and
$v_{1 \sigma} = 0$.
These scatterings, whichever is magnetic or nonmagnetic, contribute to \Tc~enhancement in the
$s_\pm$-wave state.
As discussed in \S3.3, whether \Tc~is enhanced or suppressed by the interband scattering for the
$s_\pm$ wave and also for the $s_{++}$ wave depends on the details of electron transfer,
$u_{i \sigma}$ and $v_{i \sigma}$, between the $f$-orbitals and conduction bands at an impurity.
As in eq.~(\ref{eqn:Hex}), we introduce the impurity interaction Hamiltonian for the singlet-triplet
configuration for the $\Gamma$ ($= \Gamma_4, \Gamma_5$) scattering,
\begin{align}
& \H'_\Gamma = \sum_{X = S,Q} \sum_{\bsR_\gamma} \sum_{mn}
         \int\rd\br a_{\gamma m}^\dagger a_{\gamma n} \delta(\br-\bR_\gamma) \cr
&~~~~~~~~~~~~ \times \bPsi^\dagger(\br) J_X^\Gamma \left[ X_{z,mn}^\Gamma X_z^{{\rm c},\Gamma}
\right. \cr
&~~~~~~~~~~~~\left.                 
                 + \frac{1}{2} \left( X_{+,mn}^\Gamma X_-^{{\rm c},\Gamma}
                 + X_{-, mn}^\Gamma X_+^{{\rm c},\Gamma} \right) \right] \bPsi(\br), \cr
&
\label{eqn:Hex2}
\end{align}
where the band electron scattering operators
\begin{align}
S^{{\rm c},\Gamma} = s^\Gamma,~~Q^{{\rm c},\Gamma} = q^\Gamma,
\end{align}
are rewritten on the basis of the eight-dimensional vector $\bPsi$ in eq.~(\ref{eqn:Psi8}).
\par


\end{document}